\documentclass{aa}  

\usepackage{graphicx}
\usepackage{txfonts}
\usepackage[colorlinks,citecolor=blue]{hyperref}
\begin{document} 

\title{A global model of particle acceleration at pulsar wind termination shocks}

\author{Beno\^it Cerutti \inst{1}\and Gwenael Giacinti\inst{2}}

\institute{Univ. Grenoble Alpes, CNRS, IPAG, 38000 Grenoble, France\\
           \email{benoit.cerutti@univ-grenoble-alpes.fr}
           \and
           Max-Planck-Institut f\"ur Kernphysik, Postfach 103980, 69029 Heidelberg, Germany\\
           \email{gwenael.giacinti@mpi-hd.mpg.de}
           }

\date{Received \today; accepted \today}

 
\abstract
{Pulsar wind nebulae are efficient particle accelerators, and yet the processes at work remain elusive. Self-generated, microturbulence is too weak in relativistic magnetized shocks to accelerate particles over a wide energy range, suggesting that the global dynamics of the nebula may be involved in the acceleration process instead.}
{In this work, we study the role played by the large-scale anisotropy of the transverse magnetic field profile on the shock dynamics.}
{We performed large two-dimensional particle-in-cell simulations for a wide range of upstream plasma magnetizations, from weakly magnetized to strongly magnetized pulsar winds.}
{The magnetic field anisotropy leads to a dramatically different structure of the shock front and downstream flow. A large-scale velocity shear and current sheets form in the equatorial regions and at the poles, where they drive strong plasma turbulence via Kelvin-Helmholtz vortices and kinks. The mixing of current sheets in the downstream flow leads to efficient nonthermal particle acceleration. The power-law spectrum hardens with increasing magnetization, akin to those found in relativistic reconnection and kinetic turbulence studies. The high end of the spectrum is composed of particles surfing on the wake produced by elongated spearhead-shaped cavities forming at the shock front and piercing through the upstream flow. These particles are efficiently accelerated via the shear-flow acceleration mechanism near the Bohm limit.}
{Magnetized relativistic shocks are very efficient particle accelerators. Capturing the global dynamics of the downstream flow is crucial to understanding them, and therefore local plane parallel studies may not be appropriate for pulsar wind nebulae and possibly other astrophysical relativistic magnetized shocks. A natural outcome of such shocks is a variable and Doppler-boosted synchrotron emission at the high end of the spectrum originating from the shock-front cavities, reminiscent of the mysterious Crab Nebula gamma-ray flares.}

               
\maketitle


\section{Introduction}

Pulsar wind nebulae are archetypal cosmic particle accelerators. The most studied amongst them, the Crab Nebula, presents one of the best known examples of a purely nonthermal emission spectrum extending over 20 orders of magnitude in frequency range, from 100~MHz radio waves to 100~TeV gamma rays \citep{2010A&A...523A...2M}. The bulk of the emission is almost certainly of synchrotron origin and extends up to the synchrotron burnoff limit, namely $\lesssim 100~$MeV \citep{1996ApJ...457..253D}, and slightly beyond during gamma-ray flares \citep{2011Sci...331..739A, 2011Sci...331..736T}. The electron spectrum is a broad power-law distribution spreading over at least eight orders of magnitude in particle Lorentz factor, $10\lesssim\gamma\lesssim 10^9$, with a major spectral break about half way, $\gamma\sim 10^5$. Below this break, the power law is hard and is responsible for the radio to infrared emission. Above the break, the spectrum steepens significantly and forms the optical to 100 MeV emission. Within the classical models of \citet{1974MNRAS.167....1R} and \citet{1984ApJ...283..710K}, the break as well as the high-energy component are interpreted as the injection of electron--positron pairs by the pulsar wind which are then accelerated at the wind termination shock front. This scenario is all the more promising as the slope of the injected particles above the break coincides with the first-order Fermi acceleration prediction, that is, $dN/d\gamma\propto\gamma^{-2.2}$ \citep{1998PhRvL..80.3911B, 2000ApJ...542..235K, 2001MNRAS.328..393A, 2017SSRv..207..319P}.

Nevertheless, particle acceleration is dramatically suppressed in the presence of a mean magnetic field transverse to the shock normal \citep{1988PhRvL..61..779L, 1990ApJ...353...66B, 1992ApJ...391...73G}, as found in pulsar wind nebulae where the magnetic field structure is mostly toroidal. If the plasma magnetization parameter $\sigma$, defined as the magnetic to particle enthalpy density ratio, is $\sigma\gtrsim 10^{-2}$, particles are unable to return to the shock front. Therefore, plasma turbulence is too weak to scatter particles back and forth multiple times across the shock as needed for the first-order Fermi process to operate \citep{2010MNRAS.402..321L, 2013ApJ...771...54S, 2018MNRAS.477.5238P}. Global three-dimensional (3D) magnetohydrodynamic (MHD) simulations of the Crab Nebula favor a mean plasma magnetization of order unity which can locally reach up to $\sigma\approx 10$ at high latitudes \citep{2014MNRAS.438..278P}. Thus, Fermi acceleration should be quenched, while at the same time these simulations indicate that particle acceleration most likely occurs within the equatorial regions of the shock front \citep{2014MNRAS.438..278P, 2015MNRAS.449.3149O}.

The conclusion that magnetized relativistic shocks do not accelerate particles implicitly relies on the assumption that plasma turbulence must be self-generated within the flow, like in unmagnetized shocks where the Weibel instability seeds plasma turbulence \citep{1999ApJ...526..697M, 2008ApJ...682L...5S}. This may not be the case and this is why recent studies have looked for an externally driven source of plasma turbulence, such as for example corrugations in the shock front \citep{2016JPlPh..82d6301L, 2018MNRAS.475.2713D} or the large-scale dynamics of the nebula driven by magnetic pitching and current-driven instabilities \citep{1998ApJ...493..291B, 2014MNRAS.438..278P}. Another possible solution is to consider the equatorial belt where the magnetic field vanishes by symmetry, and therefore where the Fermi process could operate \citep{2018ApJ...863...18G}.

The idea of driven magnetic reconnection within the large-scale pulsar wind current sheet at the shock front has also been considered to circumvent the above difficulties \citep{2003MNRAS.345..153L, 2007A&A...473..683P}, but this scenario requires an unusually high pair plasma supply \citep{2011ApJ...741...39S} and assumes that negligible dissipation took place in the current sheet before the shock, which may not happen \citep{1990ApJ...349..538C, 2017A&A...607A.134C}. Another particle acceleration mechanism involves electron acceleration by the absorption of ion cyclotron waves emitted at the shock front \citep{1992ApJ...390..454H, 2006ApJ...653..325A}, but this model requires a high injection rate of ions in the wind. All things considered, the origin of particle acceleration in pulsar wind nebulae remains elusive (see reviews by \citealt{2009ASSL..357..421K, 2020arXiv200104442A}).

In this work, we revisit the model of particle acceleration in relativistic magnetized shocks, taking into account a realistic latitudinal dependence of the transverse magnetic field at the shock front as expected from the theory of pulsar winds \citep{1973ApJ...180L.133M, 1999A&A...349.1017B}, in contrast with previous models which assume a uniform field. In essence, we extend the model proposed by \citet{2018ApJ...863...18G} to a larger latitudinal extent and use two-dimensional (2D) ab-initio particle-in-cell (PIC) simulations. Here, we focus our attention on the X-ray- and $<100$~MeV gamma-ray-emitting electrons only. The remainder of the paper is organized as follows. Section~\ref{sect_setup} describes the physical model, the numerical setup, and the list of runs performed in this study. Simulation results are presented in Sections~\ref{sect_dynamics}-\ref{sect_synchrotron}, which are further discussed in Section~\ref{sect_conclusion}, with particular emphasis on the Crab Nebula.

\section{Numerical setup}\label{sect_setup}

Our setup is inspired from previous PIC simulations of relativistic shocks \citep{2008ApJ...682L...5S, 2013ApJ...771...54S, 2018MNRAS.477.5238P}. It is a Cartesian box initially filled with an ultra-relativistic cold and magnetized beam of electron--positron pairs propagating along the $+x$-direction, which mimics the radial direction in this case. The right boundary reflects the particles and the fields with no loss of energy in order to form two counter-streaming beams, which eventually leads to the formation of the shock. The key difference with previous studies is the anisotropic transverse field profile along the shock front, here along the $y-$direction, which mimics the latitude. This new setup leads to additional numerical complications that we describe below.

\subsection{Fields}

\begin{figure}
\centering
\includegraphics[width=\hsize]{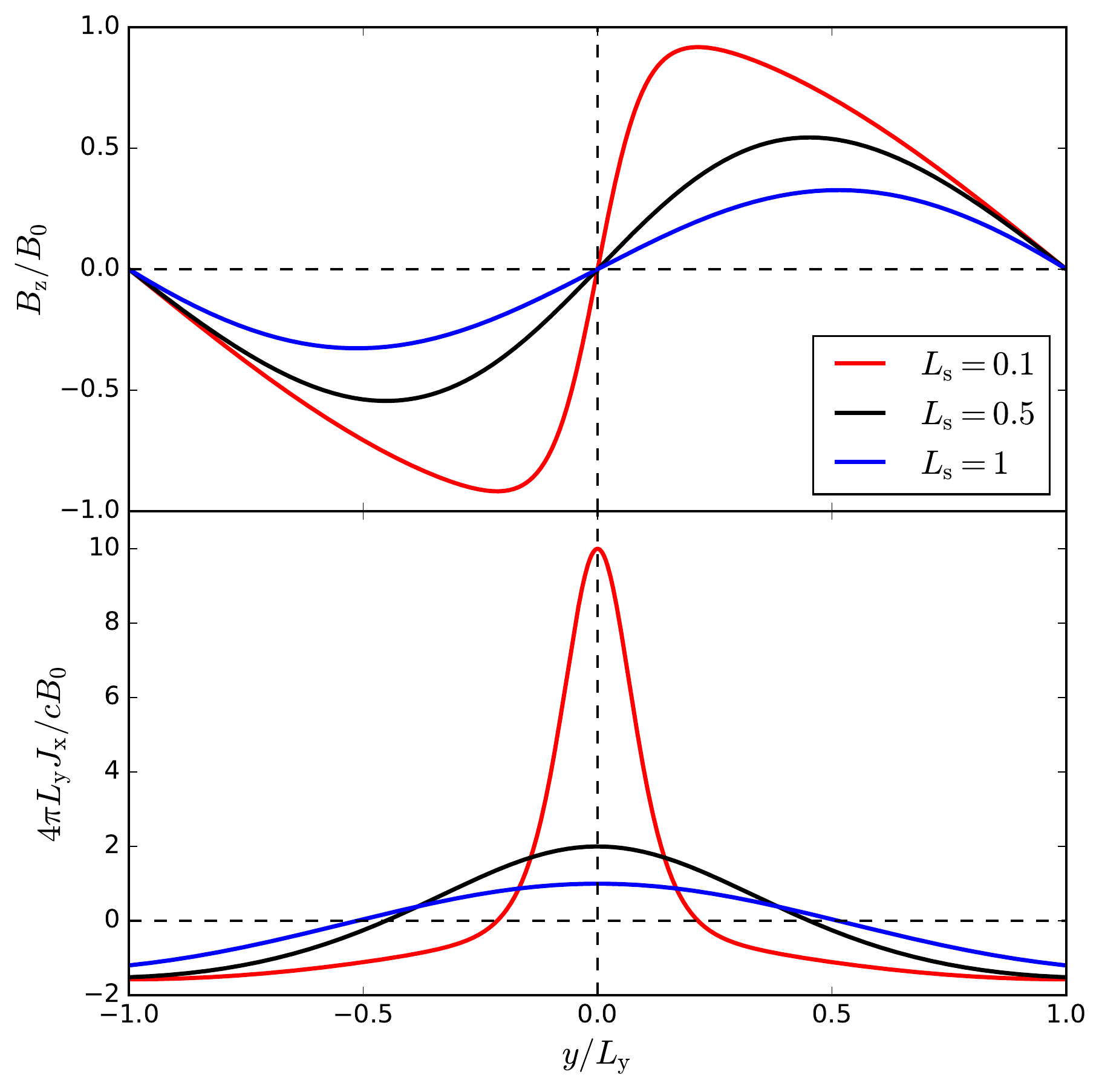}
\caption{Top: Transverse magnetic field profile in the pulsar wind along the $y$-direction, $B_{\rm z}(y)$, according to Eq.~(\ref{eq_bz}) for $L_{\rm s}/L_{\rm y}=0.1$, $0.5$, $1$. Bottom: Corresponding electric current profile, $J_{\rm x}(y)$.}
\label{fig_initial}
\end{figure}

In a split-monopole configuration \citep{1999A&A...349.1017B}, an oscillatory current sheet forms within the wind at the interface between the two magnetic polarities and fills a spherical wedge in the equatorial regions. The wind zone containing the sheet is called the striped wind \citep{1990ApJ...349..538C, 2009ASSL..357..421K}. Assuming that the sheet has fully dissipated before the wind enters the shock, we are left with the DC and axisymmetric component of the toroidal magnetic field, $B_{\phi}$, and therefore the problem which was initially three-dimensional becomes essentially two-dimensional. Translated into Cartesian coordinates and assuming that the neutron star angular velocity and magnetic field vectors fulfill $\mathbf{\Omega}\cdot\mathbf{B}>0$, a good proxy for the out-of-plane magnetic field profile is\footnote{We chose this profile over the more realistic solution given by, e.g., \citet{2013MNRAS.428.2459K}, because the derivative of the latter is discontinuous at the boundaries of the striped wind leading to spurious numerical effects we were not able to mitigate. The profile proposed here in Eq.~(\ref{eq_bz}) is a good compromise between physical realism and numerical convenience.}
\begin{equation}
B_{\rm z}(y) = B_0\tanh\left(\frac{y}{L_{\rm s}}\right)\sin\theta,
\label{eq_bz}
\end{equation}
where the $z$-direction plays the role of the toroidal direction, $B_0$ is the fiducial magnetic field strength in the upstream medium, and $L_{\rm s}$ is the spatial extent of the striped wind region set by the inclination angle between the magnetic axis and the pulsar spin axis, such that $\chi=\pi L_{\rm s}/2L_{\rm y}$. The angle $\theta=\pi(y+L_{\rm y})/2L_{\rm y}$ is the polar angle such that $y=0$ ($\theta=\pi/2$) should be understood as the equatorial plane, while $y=\pm L_{\rm y}$ should be interpreted as the poles ($\theta=0,~\pi$). These three places all have in common that the field vanishes exactly; this plays an important role in the following. Nevertheless, one should keep in mind that we neglect the curvature of the shock front with this Cartesian setup. Figure~\ref{fig_initial} (top panel) shows the dependence of $B_z$ for a nearly aligned pulsar with $L_{\rm s}/L_{\rm y}=0.1$ ($\chi=9^{\rm o}$), $L_{\rm s}/L_{\rm y}=0.5$ ($\chi=45^{\rm o}$), and for an orthogonal rotator $L_{\rm s}/L_{\rm y}=1$ $(\chi=90^{\rm o}$).

The background electric field is the ideal advection field
\begin{equation}
\mathbf{E}=-\frac{\mathbf{V}\times\mathbf{B}}{c},
\label{eq_ey}
\end{equation}
where $\mathbf{V}$ is the bulk plasma velocity, and $c$ the speed of light. For simplicity, we assume there is no latitudinal dependence of the wind velocity, $V(y)=V_0$, and so $E_{\rm y}(y)=V_0B_{\rm z}(y)/c$. The dimensions of the box in each direction are $x\in[0,L_{\rm x}]$ and $y\in[-L_{\rm y},L_{\rm y}]$. For numerical convenience, we apply periodic boundary conditions for both the fields and the particles along the $y$-directions. This choice has also a physical motivation because the toroidal field direction changes sign across the rotation axis.

\subsection{Current and charge densities}

Electron--positron pairs are continuously injected throughout the duration of the simulation. To save on computing time, new particles are uniformly created by an injector receding at the speed of light away from the right boundary, which is initially located at $x=0.95L_{\rm x}$. According to Amp\`ere's law, the pulsar wind must carry the following electric current density, $4\pi J_{\rm x}=cdB_{\rm z}/dy$, and therefore
\begin{equation}
J_{\rm x}=\frac{cB_0}{4\pi}\left[\frac{\sin\theta}{L_{\rm s}\cosh^2\left(y/L_{\rm s}\right)}+\frac{\pi}{2L_{\rm y}}\tanh\left(\frac{y}{L_{\rm s}}\right)\cos\theta\right]=J_+ +J_-.
\end{equation}
The first term, $J_+$, is strictly positive; it peaks at the equator ($y=0$) and vanishes at the poles ($y=\pm L_{\rm y}$), while the second term, $J_-$, is strictly negative with a minimum at the poles and vanishes at the equator (see the total current profile in Figure~\ref{fig_initial}, bottom panel). The total electric current passing through the $yz$-plane is zero as expected for a steady state wind. According to Gauss's law, the motion electric field in Eq.~(\ref{eq_ey}) leads to the following distribution of electric charges in the wind
\begin{equation}
\rho=\frac{1}{4\pi}\frac{dE_{\rm y}}{dy}=\frac{V_0}{c^2}J_{\rm x},
\end{equation}
so that $J_{\rm x}\approx \rho c$ in the ultrarelativistic limit. 

\subsection{Plasma density}

Assuming that both species move along the $+x$-direction at the same speed $V_0$, the minimum amount of pairs required to model both the current and the charge densities in the wind is to consider the positronic density profile
\begin{equation}
n_+=\frac{V_0 J_+}{ec^2}=n^+_0\frac{\sin\theta}{\cosh^2\left(y/L_{\rm s}\right)},
\label{eq_ni}
\end{equation}
where $n^+_0=V_0 B_0/4\pi e c L_{\rm s}$, $e$ is the electron charge, and the electronic density profile
\begin{equation}
n_-=-\frac{V_0 J_-}{ec^2}=-n^-_0\tanh\left(\frac{y}{L_{\rm s}}\right)\cos\theta,
\label{eq_ne}
\end{equation}
where $n^-_0=V_0 B_0/8 e c L_{\rm y}$. Although polarized, pulsar winds are most likely quasi-neutral, meaning that the plasma density greatly exceeds the charge density, $n\gg |\rho|/e$. Thus, on top of these minimum densities, we add a uniform neutral density of pairs $n_0$, so that $n_{\rm e}=n_0+n_-$ for the total electron density and $n_{\rm p}=n_0+n_+$ for the total positron density. In the simulations, this fiducial density is set by the chosen upstream magnetization parameter,
\begin{equation}
\sigma_0=\frac{B^2_0}{8\pi\Gamma_0 n_0 m_{\rm e} c^2},
\end{equation}
where $m_{\rm e}$ is the electron rest mass and $\Gamma_0=(1-V_0^2/c^2)^{-1/2}$ is the wind bulk Lorentz factor. Thus, the density contrast is
\begin{equation}
\frac{n^+_0}{n_0}=\frac{2V_0}{c}\frac{\sigma_0 R_0}{L_{\rm s}},
\end{equation}
\begin{equation}
\frac{n^-_0}{n_0}=\frac{\pi V_0}{c}\frac{\sigma_0 R_0}{L_{\rm y}},
\end{equation}
where
\begin{equation}
R_0=\frac{\Gamma_{\rm 0}m_{\rm e}c^2}{eB_0}
\end{equation}
is the fiducial particle Larmor radius.

\subsection{Scale separation}

The Crab Nebula spectrum suggests that the wind injects $\sim$1~TeV pairs at the shock front immersed in a $B_0\sim 200\mu$G background field. The fiducial particle Larmor radius is then of order $R_0=1~\rm{TeV}/eB_0\sim 1.6\times 10^{13}$cm. If one compares this gyroradius with the shock radius $R_{\rm sh}\sim 0.1$pc, the scale separation is $R_{\rm sh}/R_0\sim 2\times 10^4$ such that we can verify that $n_{\pm}/n_0\ll 1$ as expected even for high magnetization $\sigma_0\lesssim 100$ (except perhaps in the equatorial plane for a nearly aligned pulsar).

The main numerical challenge is to reach a sufficiently large separation of scales between the microscopic Larmor radius and plasma skin-depth scales, where particle acceleration processes take place, and the global shock size scale. The most stringent constraint in PIC simulations is to resolve the plasma skin depth $d_{\rm e}$ and plasma frequency $\omega_{\rm pe}$ scales and therefore these quantities determine the minimum spatial and time resolution of the simulations. In all runs, the fiducial skin-depth scale defined in the upstream flow is resolved by eight cells in all directions, where 
\begin{equation}
d_{\rm e}=\sqrt{\frac{\Gamma_{0} m_{\rm e}c^2}{8\pi n_{0}e^2}}=\sqrt{\sigma_0}R_0.
\end{equation}
The plasma density and the mean particle Lorentz factor in the downstream medium can differ significantly from the upstream parameters due to the compression of the flow and particle acceleration. A posteriori, we found that the plasma skin depth in the downstream flow is resolved by at least five cells in all the simulations.

The simulation time-step $\Delta t$ is determined by the usual Courant-Friedrich-Lewy condition, such that $\omega_{\rm pe}\Delta t\approx 8.75 \times 10^{-2}$. The largest simulation contains $65536\times8192$ cells along the $x$- and $y$-directions, which corresponds to a $8192 d_{\rm e}\times 1024 d_{\rm e}$ box size. In this work, we do not perform a systematic study of the effect of the transverse size of the shock in order to focus our attention on the largest possible sizes. We inject 16 particles per cell per time-step. We ran simulations for $\sigma_0=0,~0.1,~1,~10,~30,~100$, which translates into physical box sizes ranging from $2590 R_0\times 324 R_0$ for $\sigma_0=0.1$ to $44869 R_0\times 5609 R_0$ for $\sigma_0=30$, which is close to the scale separation we are seeking for the Crab Nebula. Due to the strong anisotropy of the transverse magnetic field profile, the average magnetization in the wind is $\bar{\sigma}_0\approx 0.15\sigma_0$ for $L_{\rm s}/L_{\rm y}=0.5$ ($\bar{\sigma}_0\approx 0.4\sigma_0$ for $L_{\rm s}/L_{\rm y}=0.1$ and $\bar{\sigma}_0\approx 0.065\sigma_0$ for $L_{\rm s}/L_{\rm y}=1$). The cyclotron frequency $\omega_{\rm c}\Delta t=8.75\times 10^{-2}\sqrt{\sigma_0}$ is well resolved in all runs, even for $\sigma_0=100$. The largest simulation is integrated for about $7875\omega_{\rm pe}t$, or $43133\omega_{\rm c} t$ for $\sigma_0=30$. Scaled to the Crab Nebula, this represents a total simulation time of about $260$~days which is of the order of a few times the dynamical timescale of the nebula, $t_{\rm neb}\sim R_{\rm sh}/c\sim 100$~days.

\begin{figure*}
\centering
\includegraphics[width=9cm]{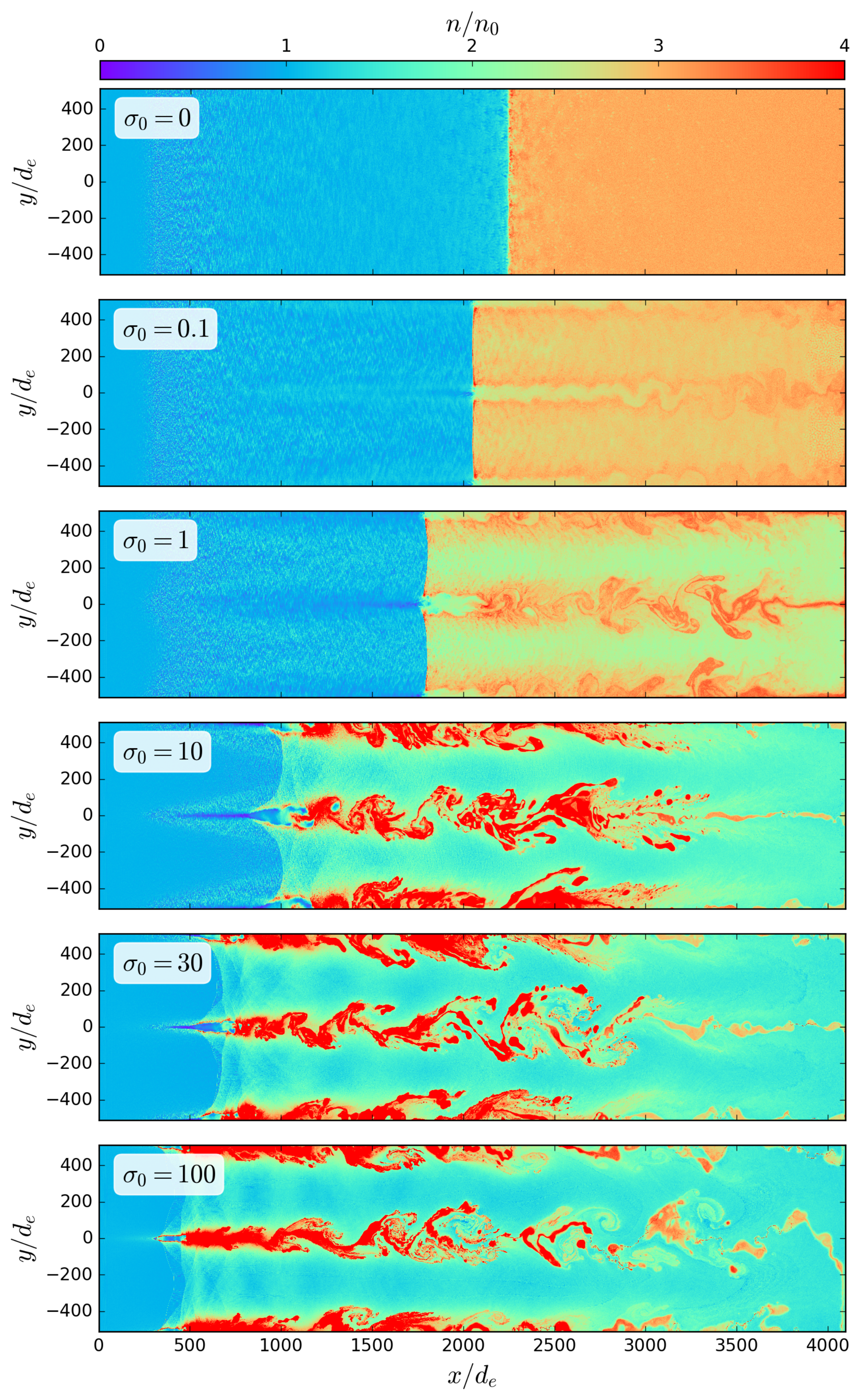}
\includegraphics[width=9cm]{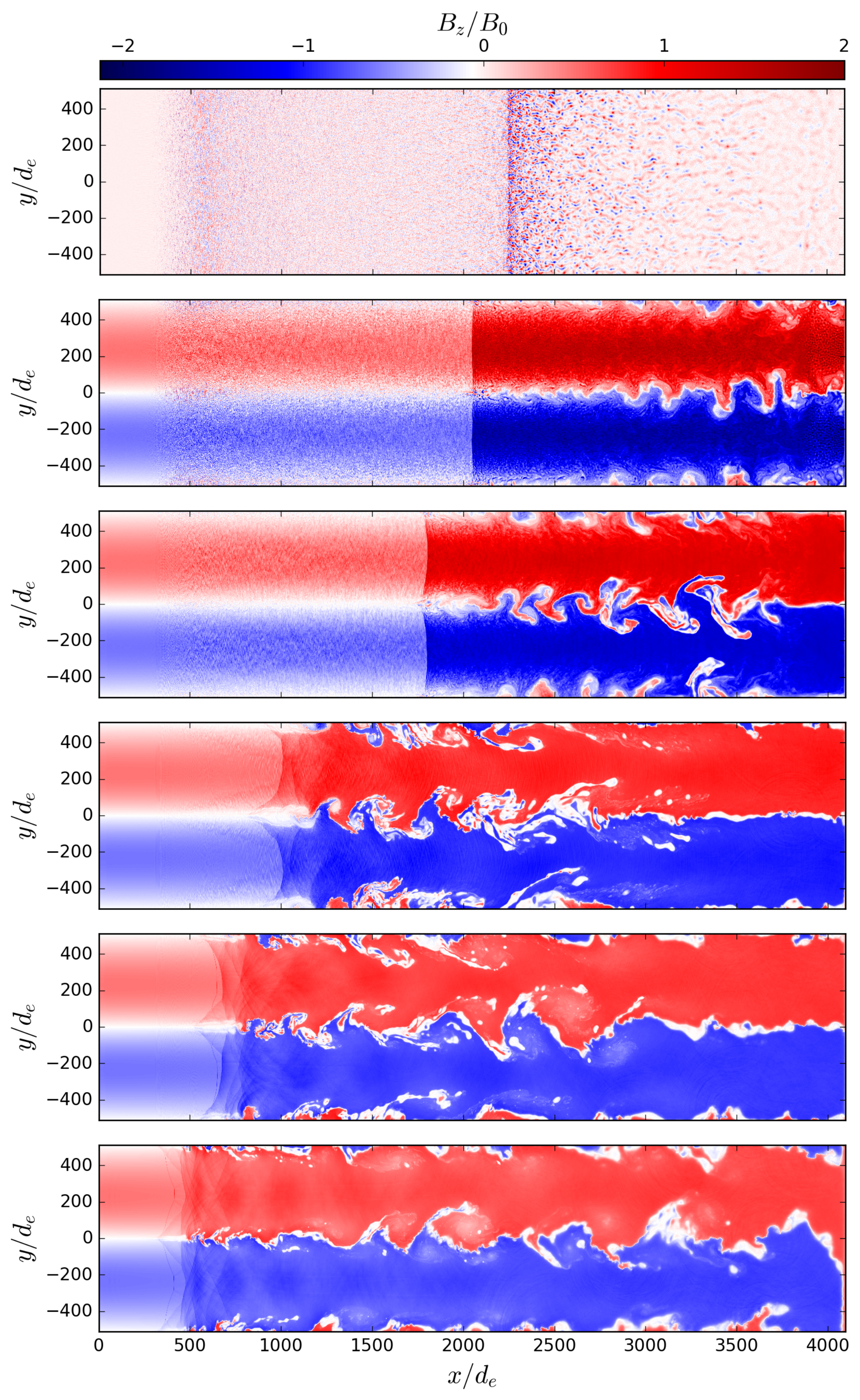}
\caption{Total plasma density (left panels) and transverse magnetic field strength (right panels) for $\sigma_0=0$, $0.1$, $1$, $10$, $30$, $100$ (from top to bottom) and $L_{\rm s}/L_{\rm y}=0.5$, at $\omega_{\rm pe}t=3920$.}
\label{fig_maps}
\end{figure*}

Aside from the separation of scale, we must also mitigate the effect of numerical Cherenkov radiation that tends to slow down and heat up ultra-relativistic beams \citep{2004JCoPh.201..665G}. This instability grows with the amplitude of the Lorentz factor of the beam. The wind Lorentz factor in pulsar winds is uncertain but is most likely very high, $\Gamma_{\rm 0}\sim 10^2-10^6$. In this work, we scale down the wind velocity to $V_{0}=0.99c$ or $\Gamma_{0}\approx 7$, which fulfills the need to have $\Gamma_{0}\gg 1$ and low numerical heating before the end of the simulation. A small temperature ($kT_{0}/m_{\rm e}c^2=10^{-2}$) is added and spatial filtering is applied to the current density to even further delay the onset of the instability. Radiative cooling (primarily synchrotron and inverse Compton) is neglected in this work.

\subsection{Summary of all runs}

All runs in this study were performed with the Cartesian version of the Zeltron PIC code \citep{2013ApJ...770..147C, 2019ascl.soft11012C}. Table~\ref{table_sim} gives the list of all runs reported in this work.

\begin{table}
\caption{\label{table_sim}List of all PIC simulations reported in this work.}
\centering
\begin{tabular}{lcccc}
\hline\hline
Run & Size (in $d_{\rm e}$) & $\sigma_0$ & $L_{\rm s}/L_{\rm y}$ & $\bar{\sigma}_0$\\
\hline
S0 & $4096\times1024$        & $0$   & $0.5$ & $0$ \\
S01 & $4096\times1024$       & $0.1$ & $0.5$ & $0.015$ \\
S1 & $4096\times1024$        & $1$   & $0.5$ & $0.15$ \\
S10 & $8192\times1024$       & $10$  & $0.5$ & $1.5$ \\
S10\_LS01 & $4096\times1024$ & $10$  & $0.1$ & $4$ \\
S10\_LS1 & $4096\times1024$  & $10$  & $1$   & $0.65$ \\
S30 & $8192\times1024$       & $30$  & $0.5$ & $4.5$ \\
S100 & $4096\times1024$      & $100$  & $0.5$ & $15$ \\
\hline
\end{tabular}
\end{table}

\begin{figure*}
\centering
\includegraphics[width=16cm]{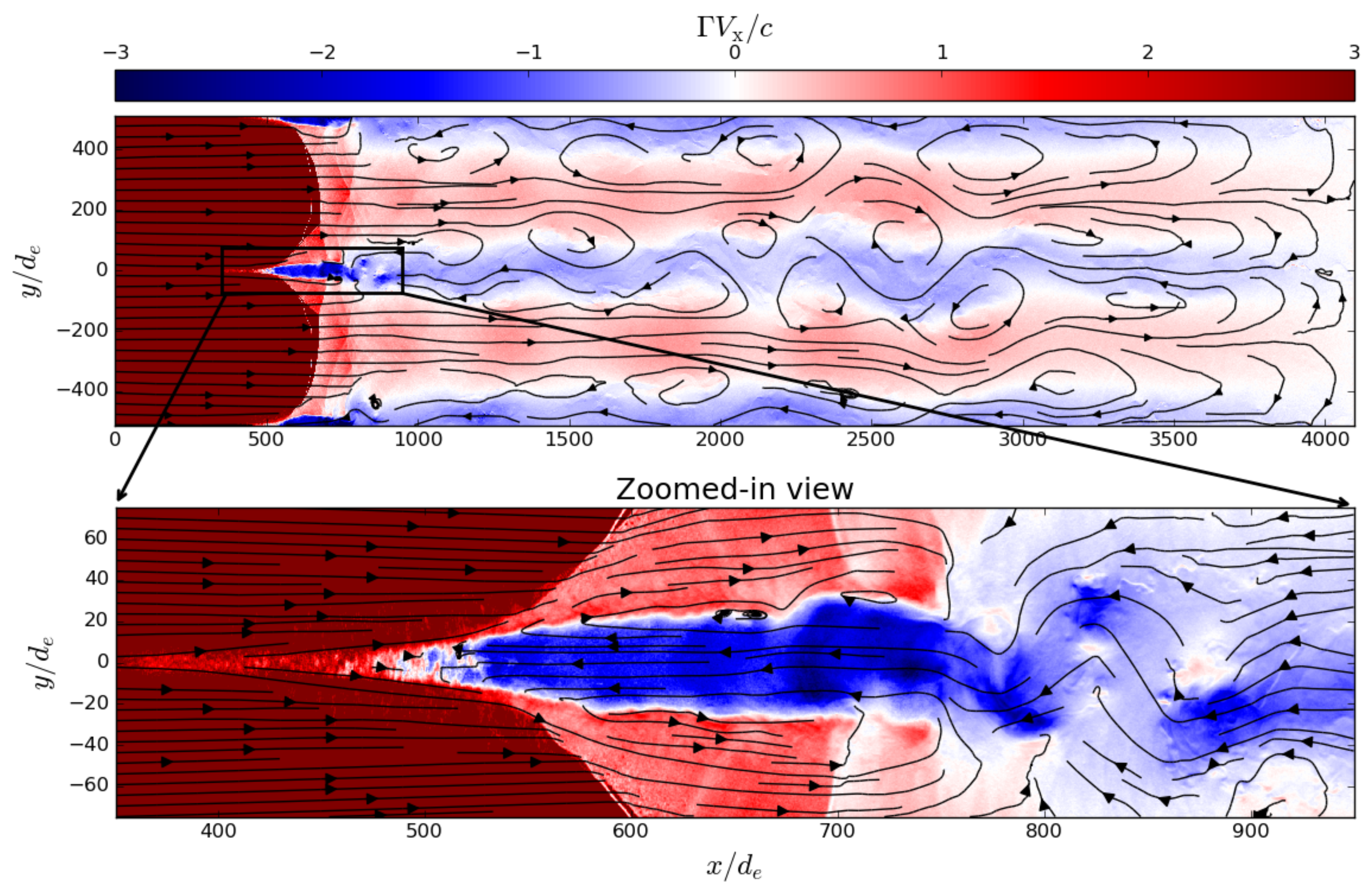}
\caption{Top: Plasma bulk momentum along the $x$-direction, $U_{\rm x}=\Gamma V_{\rm x}/c$, for $\sigma_0=30$ over the full simulation box at time $\omega_{\rm pe}t=3920$. Bottom: Zoomed-in view of the equatorial region of the shock front shown by the rectangular box in the upper panel. Solid black lines with arrows are the plasma velocity streamlines.}
\label{fig_bulk}
\end{figure*}

\begin{figure}
\centering
\includegraphics[width=9cm]{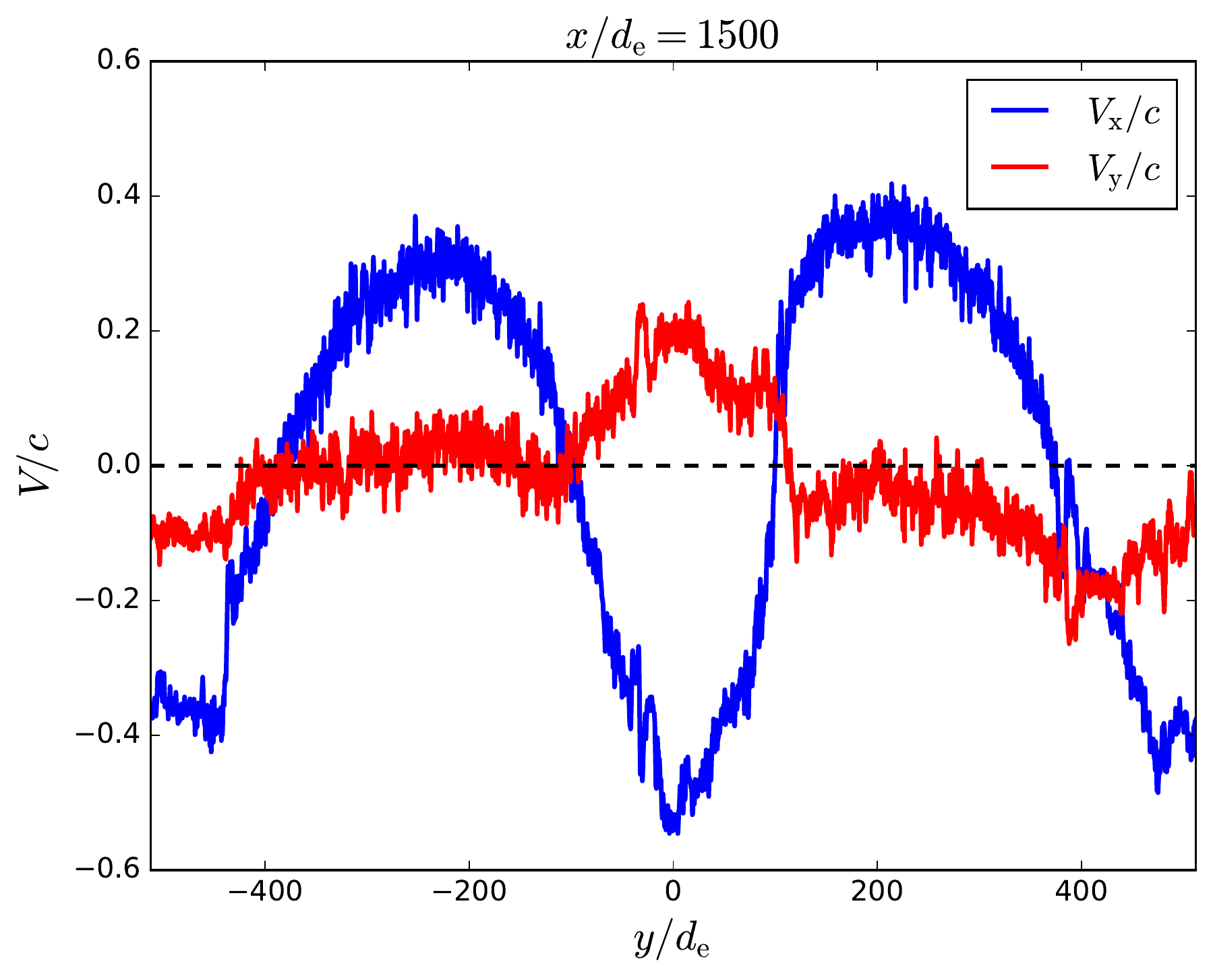}
\caption{Plasma bulk velocity profile in the downstream flow along the $x$- and $y$-directions at $x/d_{\rm e}=1500$ for $\sigma_0=30$ at time $\omega_{\rm pe}t=3920$.}
\label{fig_bulk_profile}
\end{figure}

\begin{figure*}
\centering
\includegraphics[width=\hsize]{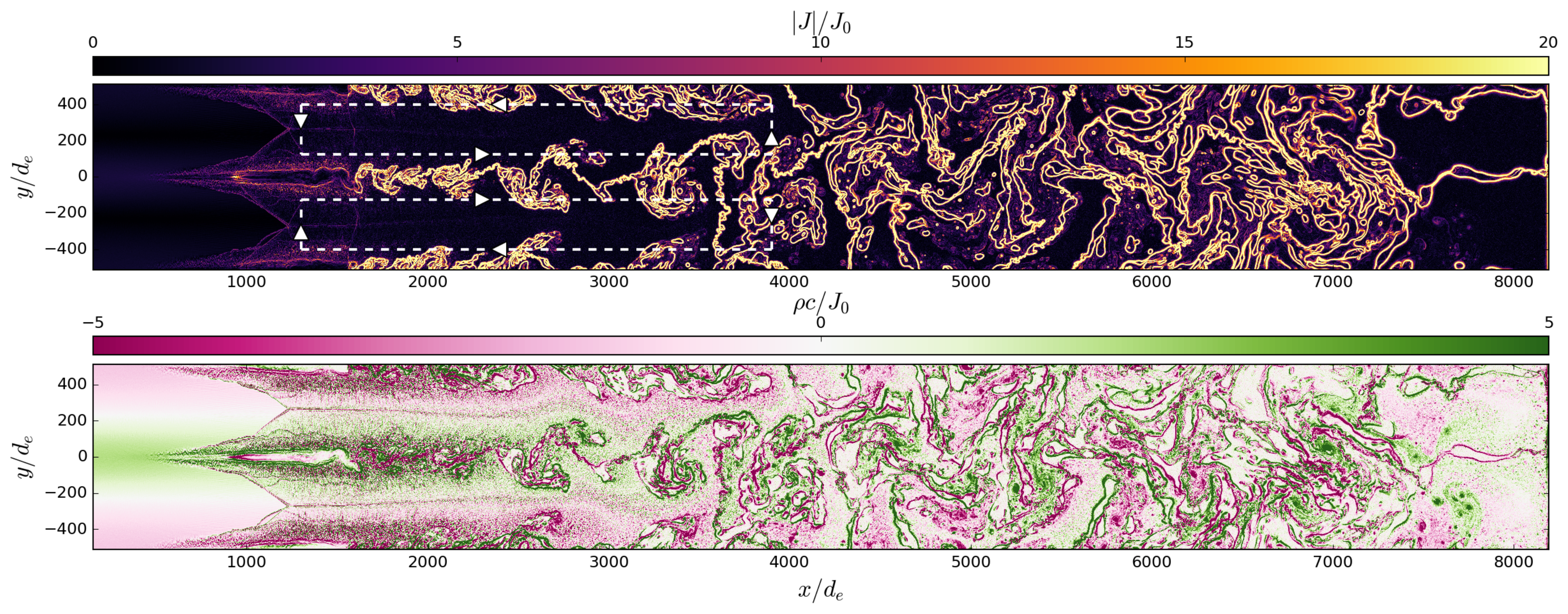}
\caption{Top: Magnitude of the total electric current density, $|J|$, normalized by the fiducial upstream current $J_0=cB_0/4\pi L_{\rm y}$ for $\sigma_0=30$ at $\omega_{\rm pe}t=7840$. The white arrows schematically show how the current flows and closes within the simulation box. Bottom: Plasma charge density, $\rho$, normalized by the fiducial upstream charge density $J_0/c$.}
\label{fig_current}
\end{figure*}

\section{Shock structure and dynamics}\label{sect_dynamics}

Figure~\ref{fig_maps} shows the shock structure at time $\omega_{\rm pe}t=3920$ in order of increasing magnetization, starting with a perfectly unmagnetized shock $\sigma_0=0$ (top panels) down to a strongly magnetized shock with $\sigma_0=100$ (bottom panel) and the transition in between these two extreme regimes. The unmagnetized case serves as a control simulation we can compare with previous studies (e.g., \citealt{2008ApJ...682L...5S, 2009ApJ...693L.127K, 2013ApJ...771...54S, 2018MNRAS.477.5238P}). As expected for an unmagnetized shock, the interaction between the incoming and reflected beams leads to the formation of Weibel filaments which then mediate magnetic turbulence and ultimately the formation of the shock. These filaments are highly visible as self-generated magnetic structures of alternative polarities, whose strength peaks at the shock front and slowly decays downstream. The filamentation proceeds in the wind ahead of the shock front due to the reflected beam of particles propagating back upstream. This region, usually called the precursor, effectively decelerates the incoming flow and sustains magnetic turbulence (see e.g., \citealt{2019PhRvL.123c5101L}), which is the key ingredient to bring particles back to the shock and initiate the Fermi process. The downstream plasma density increases by $n/n_0\approx 3$ as expected from the MHD jump conditions of an unmagnetized perpendicular shock \citep{1984ApJ...283..710K, 2018MNRAS.477.5238P}.

Adding a nonzero but subdominant field ($\sigma_0=0.1$) quenches the formation of Weibel filaments in most of the flow, except within the equatorial and polar regions where the effective magnetization is small (only a few filaments are visible in each of these regions). The upstream magnetic field is compressed downstream by a factor of approximately three as expected in the low-$\sigma$ limit. Similarly, the plasma density is compressed by approximately the same amount, with a noticeable depletion in the equatorial and polar regions. We observe a well-defined high-density rim at the shock front at intermediate heights. This is a known feature of magnetized shocks which results from the magnetic reflection of the incoming flow of particles (see, e.g., \citealt{2018MNRAS.477.5238P}). When the particles cross the shock front, they lose the support of the transverse electric field which allowed them to go in straight lines upstream. As a result, the incoming particles gyrate coherently at the shock front resulting in this characteristic plasma density bump and the emission of an electromagnetic precursor upstream \citep{1992ApJ...391...73G}.

At even higher magnetization ($\sigma_0\gtrsim 1$), the shock structure changes dramatically and we are now leaving known territory for a new phenomenology. As the magnetization increases, what appears to be a shock front travels faster upstream. In magnetized regions located at intermediate heights, the jumps in the plasma density and the magnetic field strength across the shock front also decrease with increasing magnetization. While these features are consistent with a weak shock, the equatorial and polar regions behave very differently. The flow is strongly compressed into a highly turbulent state driven by kinks (current-driven) and Kelvin-Helmholtz vortices (shear-driven). The compression of the flow into low-field regions is the result of the magnetic pressure force. The global bulk flow then quickly converges towards the pattern shown in the upper panel in Figure~\ref{fig_bulk} for $\sigma_0=30$. This figure shows the dimensionless bulk momentum of the flow along the $x$-direction, $U_{\rm x}=\Gamma V_{\rm x}/c$, and plasma velocity streamlines. In magnetized regions ($y/d_{\rm e}=\pm 256$), the flow decelerates down to about $\sim+0.3c$. In the low-field regions, the flow velocity returns back to the shock with a net bulk velocity $\approx -0.5c$ in the downstream medium. This is a major difference with uniform shocks where the plasma is at rest in the downstream region. In the transition region ($y/d_{\rm e}\sim \pm 100$), there is a strong velocity shear $\Delta V\sim 0.8 c$ (Figure~\ref{fig_bulk_profile}) which drives the formation of Kelvin-Helmholtz vortices well visible in the density maps as well as in the streamline pattern.

An intriguing and robust feature of the shock is the spearhead-shaped structures developing at the shock front in the equatorial and polar regions, which are elongated along the direction of the flow. These structures are low-field, low-density regions characterized by a mildly relativistic backflow motion up to $U_{\rm x}\approx-4$ at $x/d_{\rm e}\approx 700$ for $\sigma_0=30$ (see zoomed-in view in the bottom panel in Figure~\ref{fig_bulk}). They are also characterized by a large and abrupt velocity shear at their boundaries. The sheath-like structures around them gradually deflect the incoming flow sideways, such that there is no clear sign of a standard shock pattern here. Away from this triangular-shaped precursor drilling through the upstream medium, the incoming flow is perfectly laminar with no sign of plasma turbulence. The size of these cavities continuously grows with time without any sign of saturation. The kink-like motion of the plasma concentrated in the midplane seems to depart from the base of these structures. The plasma carries away the electric current within high-density filaments, simply referred to as `current layers' in the following. The current flows along the equator to sustain the jump in the magnetic field polarity. In the early phases of the simulation ($\omega_{\rm pe}t\lesssim 4000$), the current then flows along the $\pm y$-directions at the $x=L_{\rm x}$ boundary and reaches the polar regions where it flows in the opposite direction. This electric circuit gradually closes through the shock front and the spearhead cavities. At later stages ($\omega_{\rm pe}t\gtrsim 4000$), Kelvin-Helmholtz vortices combined with the kink lead to an efficient mixing of the downstream flow into a highly turbulent state. The top panel in Figure~\ref{fig_current} shows the total current and its schematic path within the numerical box at a late evolutionary stage ($\omega_{\rm pe}t=7840$), when a turbulent mixing state has been reached far downstream. It is important to notice that the downstream flow, and in particular the current layers and their associated cavities, are electrically charged with a net negative charge in the equator and a net positive charge at the poles (and vice versa if $\mathbf{\Omega}\cdot\mathbf{B}<0$). Before they mix and reconnect far downstream, each layer is surrounded by a low-density background plasma with the opposite sign of charge, but of the same sign as the upstream flow (Figure~\ref{fig_current}, bottom panel).

\section{Particle acceleration}

\subsection{Total spectra and maximum energy}

\begin{figure}
\centering
\includegraphics[width=\hsize]{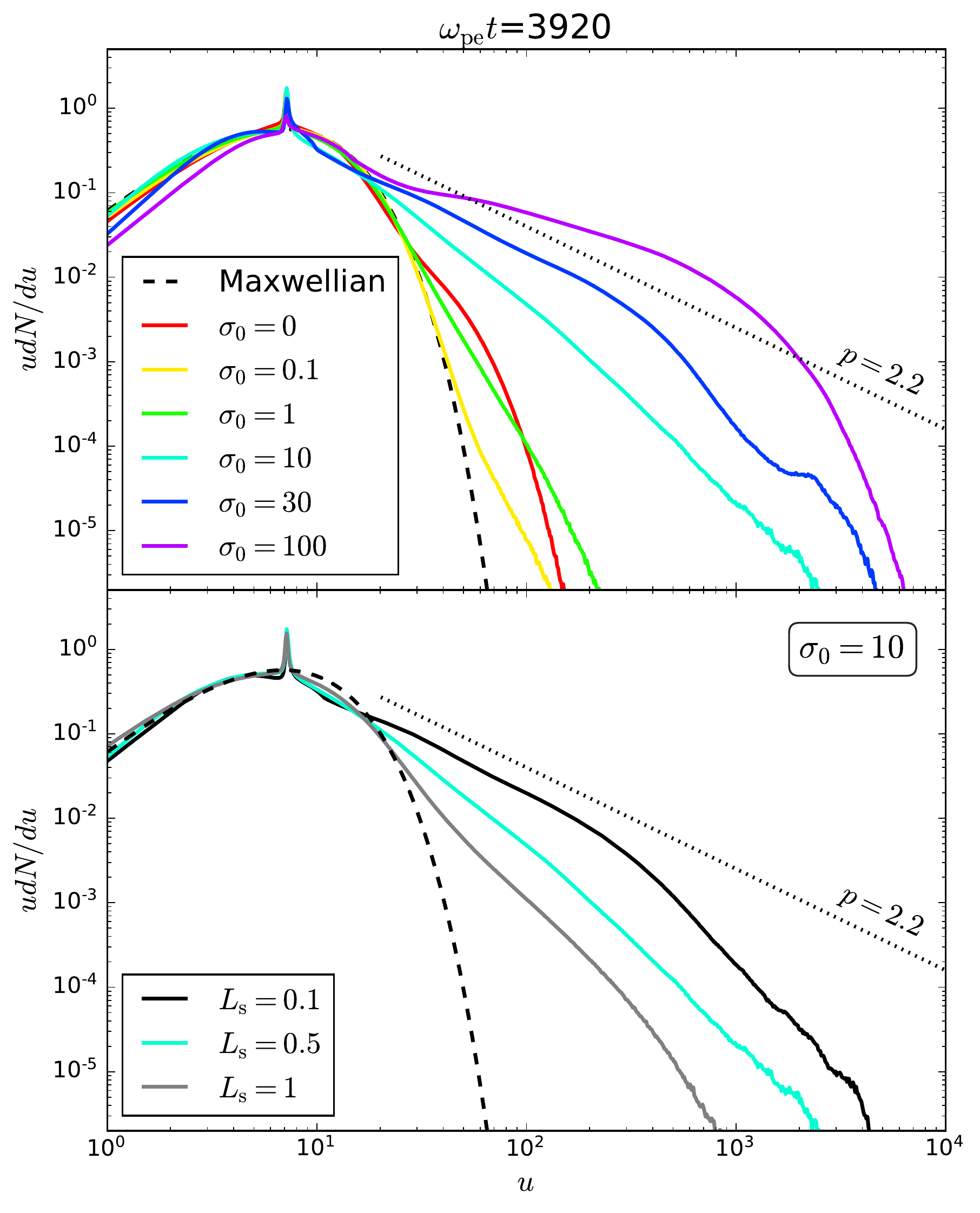}
\caption{Top: Total particle spectrum as a function of the dimensionless total particle momentum, $u=\gamma\beta$ for all the magnetizations explored in this study and at the same time $\omega_{\rm pe}t=3920$. Bottom: Effect of the striped wind filling factor on the final particle spectrum, $L_{\rm s}/L_{\rm y}$ for $\sigma_0=10$ at the same time as in the top panel. The black dotted line shows a power law $dN/du\propto u^{-p}$ of index $2.2$. The black dashed line is a 2D relativistic Maxwellian distribution for comparison.}
\label{fig_spectra}
\end{figure}

\begin{figure}
\centering
\includegraphics[width=9cm]{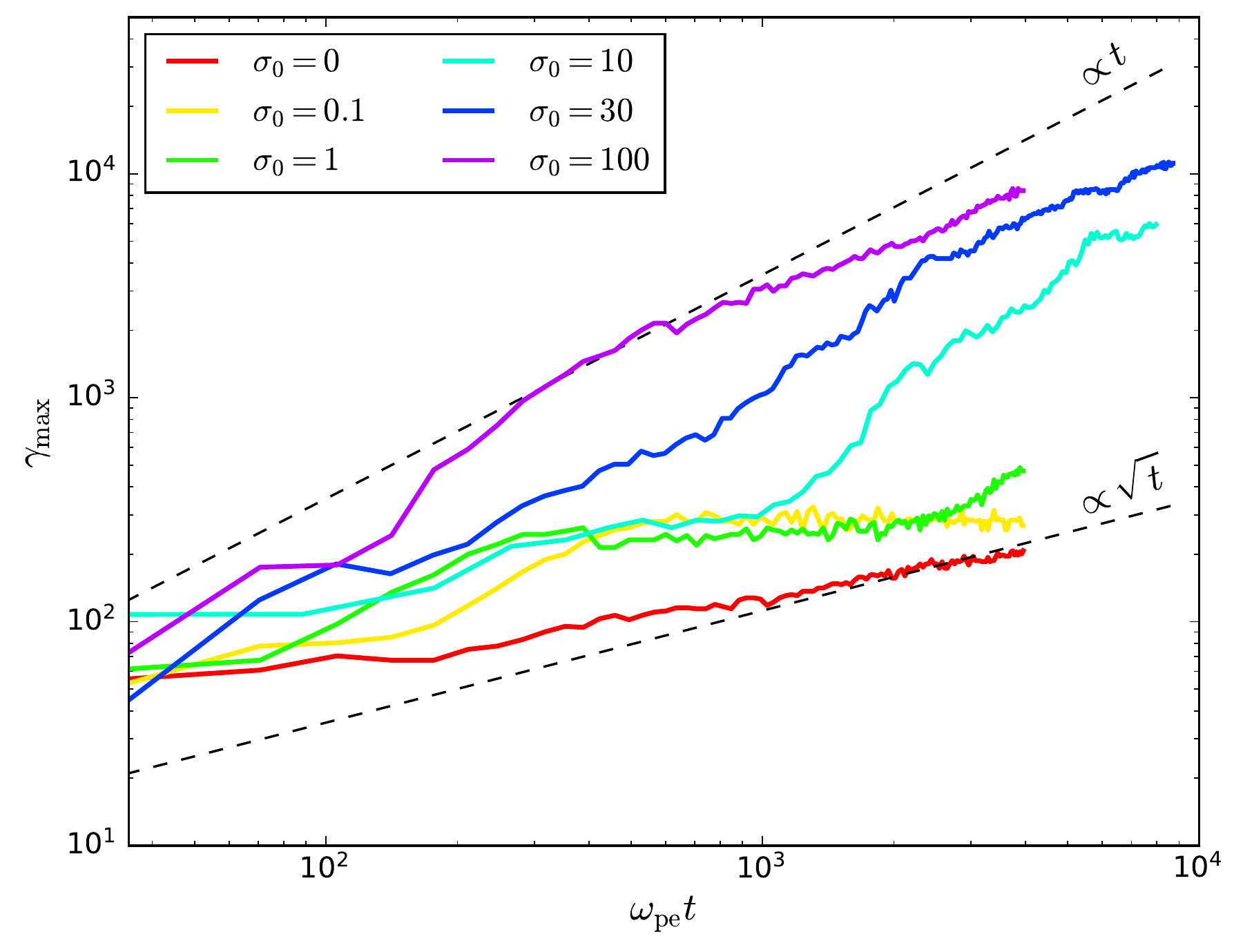}
\caption{Time evolution of the maximum particle Lorentz factor of the total spectrum, $\gamma_{\rm max}$, in all the runs. The dashed black lines show $\gamma_{\rm max}=0.5\Gamma_0\sqrt{\omega_{\rm pe}t}$ and $\gamma_{\rm max}=0.5\Gamma_0 \omega_{\rm pe}t$.}
\label{fig_gmax}
\end{figure}

\begin{figure*}
\centering
\includegraphics[width=\hsize]{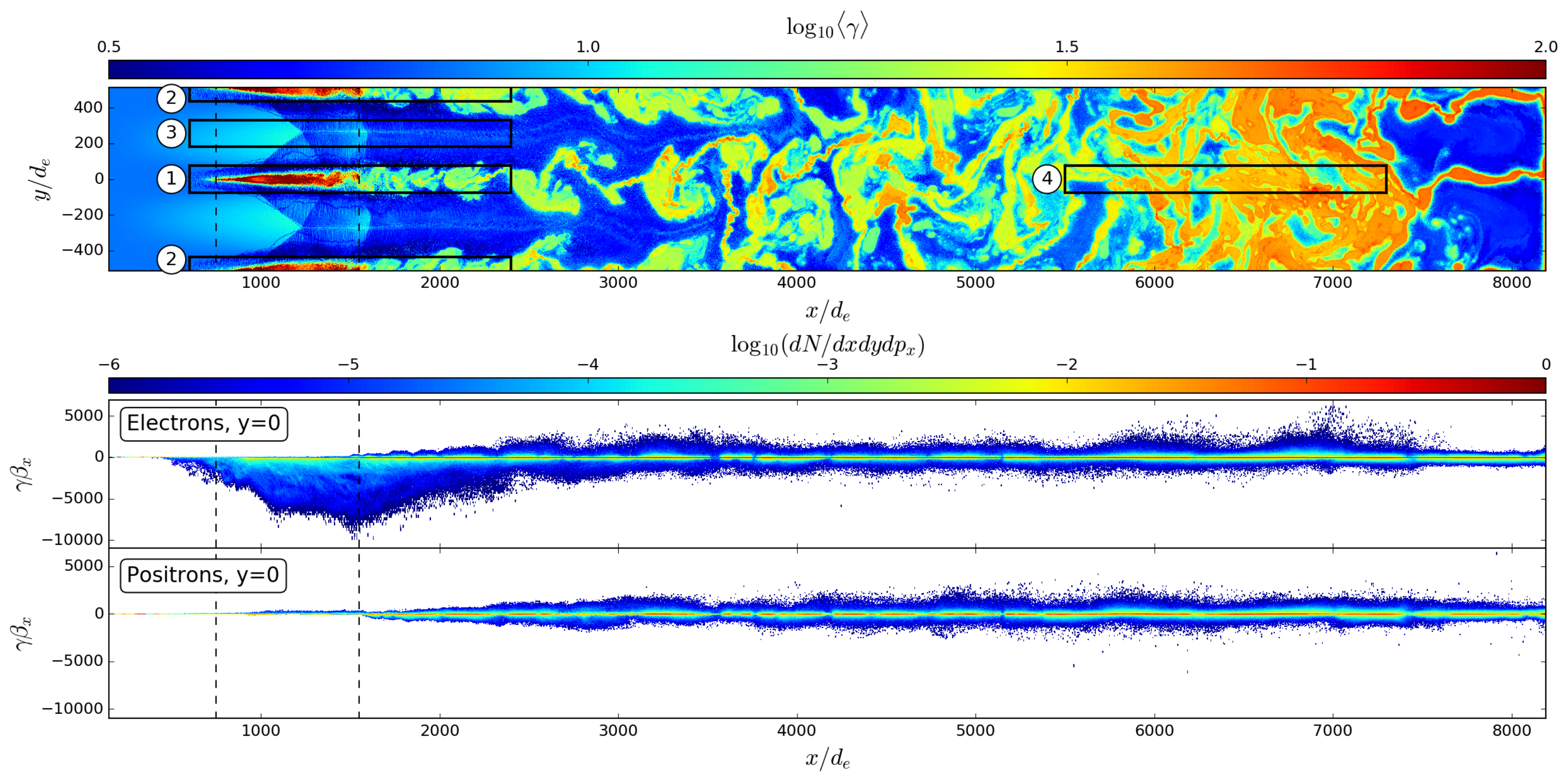}
\includegraphics[width=14cm]{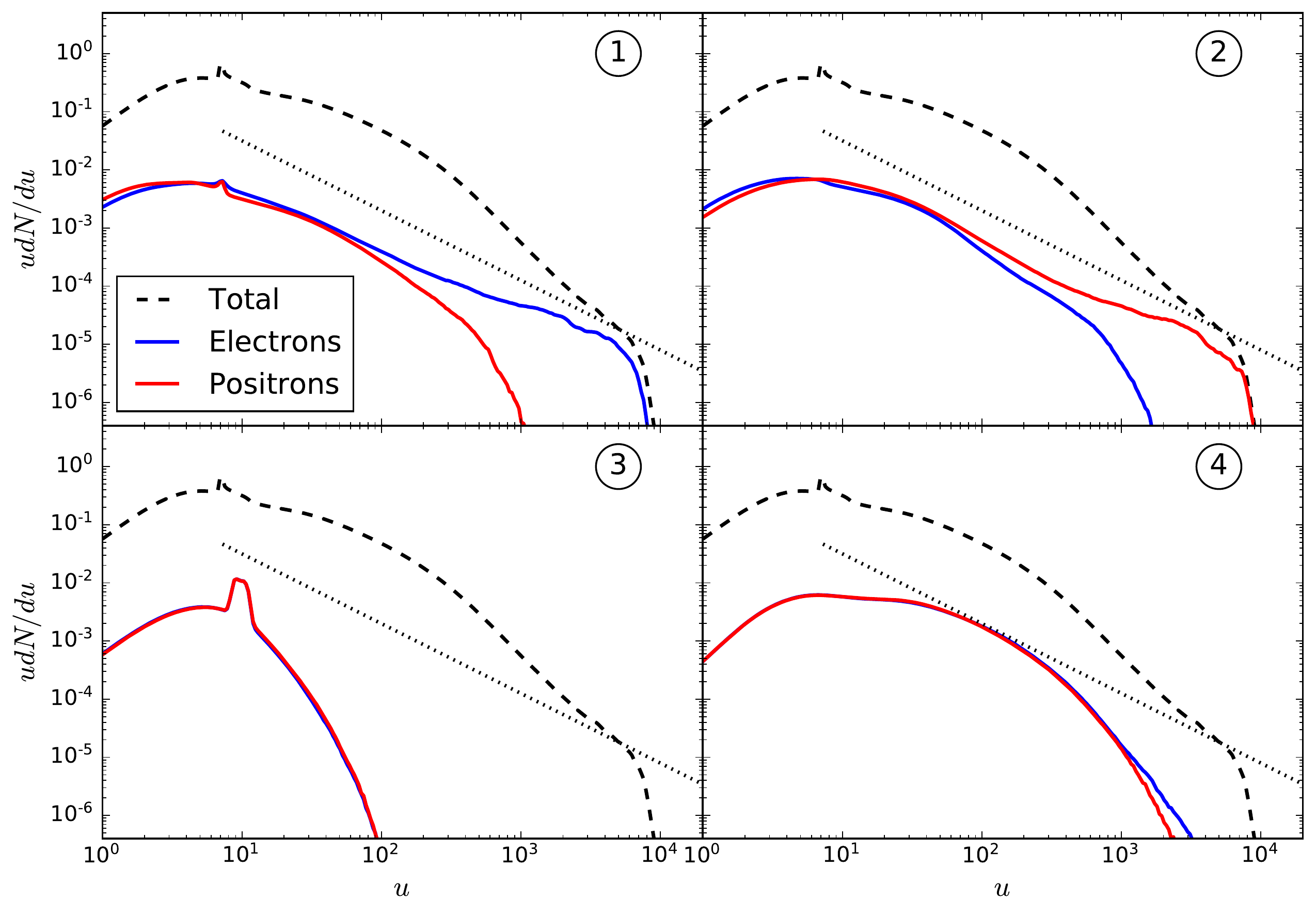}
\caption{Top panel: Mean particle Lorentz factor at $\omega_{\rm pe}t=7840$ for $\sigma_0=30$. Middle panels: Electron and positron $x$-$p_{\rm x}$ phase-space distribution $dN/dxdydp_{\rm x}$ along the midplane integrated within $y/d_{\rm e}\pm 75$. Bottom panels: Local particle spectra measured at four locations shown in the top panel and labeled from \raisebox{.5pt}{\textcircled{\raisebox{-.9pt} {1}}} to \raisebox{.5pt}{\textcircled{\raisebox{-.9pt} {4}}}. The dashed line is the total spectrum, and the dotted line is $dN/du\propto u^{-2.2}$ for reference.}
\label{fig_phase_space}
\end{figure*}

This rather complex shock structure leads to efficient nonthermal particle acceleration. The upper panel in Figure~\ref{fig_spectra} shows the total particle spectrum, $udN/du$, where $u=\gamma\beta$ is the dimensionless particle momentum at time $\omega_{\rm pe}t=3920$ for all the magnetizations simulated here. As expected, the unmagnetized shock produces a high-energy power-law tail extending beyond the thermal bath. For $\sigma_0=0.1$, particle acceleration is quenched in most of the shock front (thermal spectrum), except in the low-field regions which produces a weak excess starting below $udN/du\lesssim 10^{-4}$ and extending to the same maximum energy as in the unmagnetized shock. In contrast, strongly magnetized shocks ($\sigma_0> 1$) present a pronounced high-energy power-law tail extending to energies far beyond unmagnetized shocks, with a maximum Lorentz factor $\gamma_{\rm max}/\Gamma_{0}\sim 500$ for $\sigma_0=30$ compared with $\gamma_{\rm max}/\Gamma_{0}\sim 20$ for $\sigma_0=0$. The power law hardens as well with increasing magnetization, approaching the canonical first-order Fermi acceleration spectrum $dN/du\propto u^{-2.2}$ for $\sigma_0=30$. At this magnetization, the spectrum breaks and steepens at high energy ($\gamma\gtrsim 300$) and a new component emerges whose nature will become clear later. The bottom panel in Figure~\ref{fig_spectra} shows the dependence of the particle spectrum with the striped wind filling factor, $L_{\rm s}/L_{\rm y}$. Particle acceleration is more pronounced at low $L_{\rm s}$, which is consistent with the trend reported earlier that particle acceleration is more effective when the wind is more magnetized on average; that is, $\bar{\sigma}_0\approx 4.5$ for $L_{\rm s}/L_{\rm y}=0.1$, as compared with $\bar{\sigma}_0\approx 0.65$ for $L_{\rm s}/L_{\rm y}=1$.

Figure~\ref{fig_gmax} shows the time evolution of the maximum particle Lorentz factor of the total spectrum, $\gamma_{\rm max}(t)$. For the unmagnetized shock ($\sigma_0=0$), the maximum energy approximately grows as $\gamma_{\rm max}/\Gamma_0\approx 0.5 \sqrt{\omega_{\rm pe}t}$ without any sign of saturation, in agreement with \citet{2013ApJ...771...54S}. The square-root dependence reflects the microscopic nature of the Weibel-driven turbulence. Finite but mildly magnetized solutions ($\sigma_0=0.1,~1,~10$) show a similar acceleration rate at the early stages, but this is followed by a saturation at $\omega_{\rm pe}t\gtrsim 500$ which is particularly visible for the $\sigma_0=0.1$ solution. This saturation is related to the finite size of the turbulent region in the upstream flow, which approximately scales as the particle Larmor radius in the background field \citep{2010MNRAS.402..321L, 2013ApJ...771...54S, 2018MNRAS.477.5238P}.

For uniform shocks, there would be no more evolution of the particle spectrum. In contrast, we observe here that the maximum energy then increases again and at a much faster rate in the late evolution of the simulation. For $\sigma_0=1$, $\gamma_{\rm max}$ increases again at $\omega_{\rm pe}t\gtrsim 2\times 10^3$. For $\sigma_0=10$, it occurs earlier, at $\omega_{\rm pe}t\gtrsim 10^3$, followed by a quasi-linear evolution $\gamma_{\rm max}\propto t$. More highly magnetized solutions ($\sigma_0=30,~100$) only show a linear evolution of the maximum particle energy with time from the beginning of the simulations. The longest runs ($\sigma_0=10$ and $30$) show no sign of saturation. The linear increase of the particle energy with time is evidence for efficient particle acceleration, compatible with the Bohm regime but in sharp contrast with shock acceleration mediated by self-generated microturbulence \citep{2013ApJ...771...54S}.

\subsection{Phase space and local spectra}\label{sect_phase_space}

To gain further physical insight into the origin and location of particle acceleration, we compute the mean particle Lorentz factor, $\langle\gamma\rangle$, in each cell of the simulation as reported in the top panel in Figure~\ref{fig_phase_space} at $\omega_{\rm pe}t=7840$. High-energy particles are located in high-density regions within the current layers and inside the spearhead-shaped cavities at the shock front. A look into the $x$-$p_{\rm x}$ phase-space distribution within the midplane reveals that the latter contain the highest energy particles in the simulations, $\gamma\sim 10^4$, in the form of a strongly beamed distribution traveling against the incoming flow. Electrons (positrons) are preferentially accelerated in the equatorial (polar) cavities and vice versa if $\mathbf{\Omega}\cdot\mathbf{B}<0$. The asymmetry between both species as well as the anisotropy of their momentum distribution decrease downstream and even fully disappear where the flow becomes turbulent ($x/d_{\rm e}\gtrsim 4000$). The bottom panels in Figure~\ref{fig_phase_space} show the particle spectra measured in four areas defined in the top panel. Areas 1 and 2 are restricted to the shock-front cavities. The asymmetric acceleration between electrons and positrons is clearly visible here. In contrast to the total spectrum, the spectrum measured in these cavities is consistent with a single power law extending from $\gamma_{\rm min}=\Gamma_0$ to $\gamma_{\max}\approx 10^4$ with an index close to but slightly harder than $-2.2$. Area 3 is limited to the shock front at intermediate latitudes where particle acceleration is quenched. Area 4 focuses on the far-downstream region where the current layers reconnect and merge in a turbulent manner. The spectrum is hard at low energies, but cuts off noticeably below the maximum energy measured in the cavities. This difference explains the high-energy break at $\gamma\gtrsim 300$ in the total spectrum, beyond which the spectral component from the shock-front cavities takes over.

\subsection{Particle trajectories}

\begin{figure*}
\centering
\includegraphics[width=9cm]{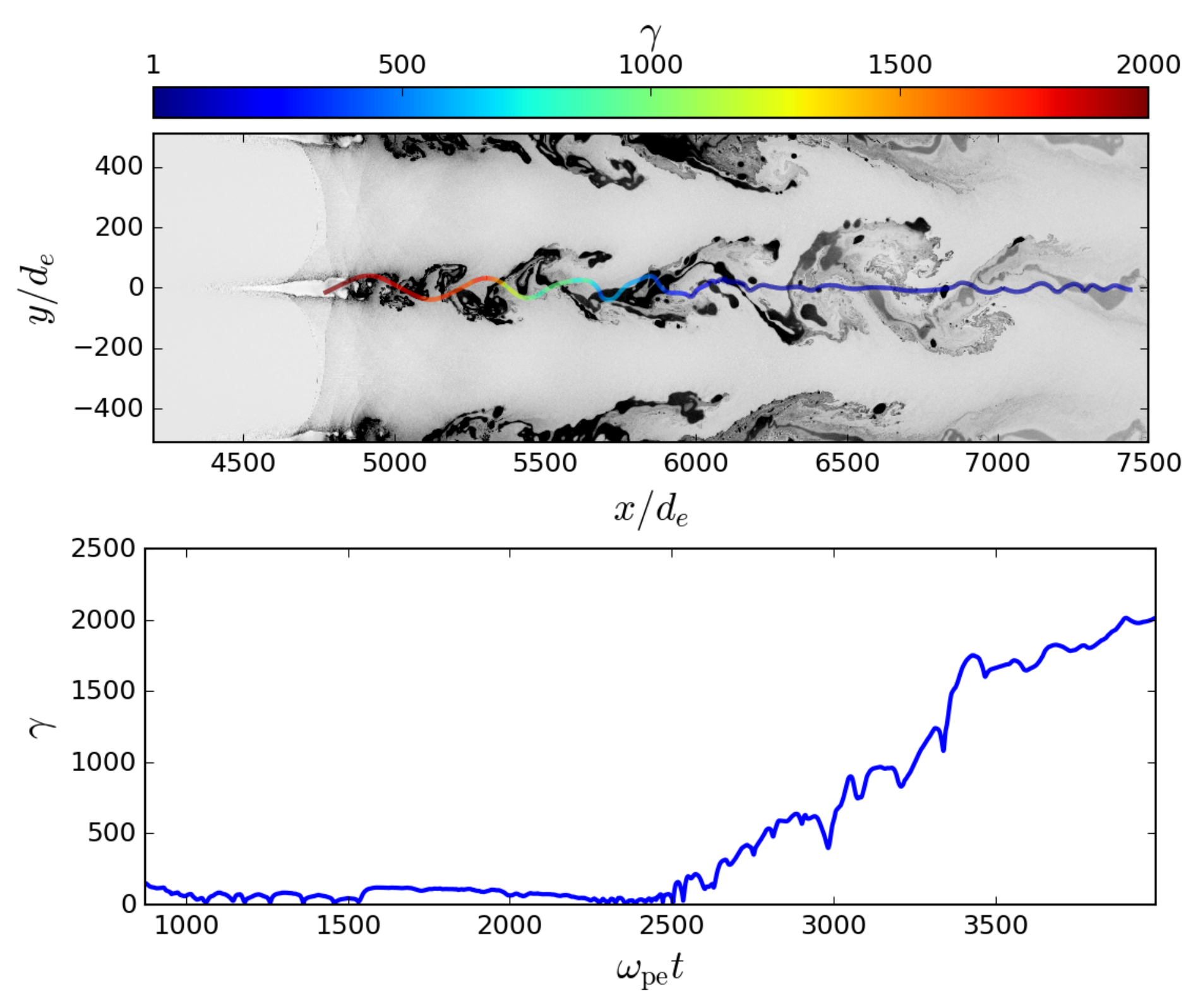}
\includegraphics[width=9cm]{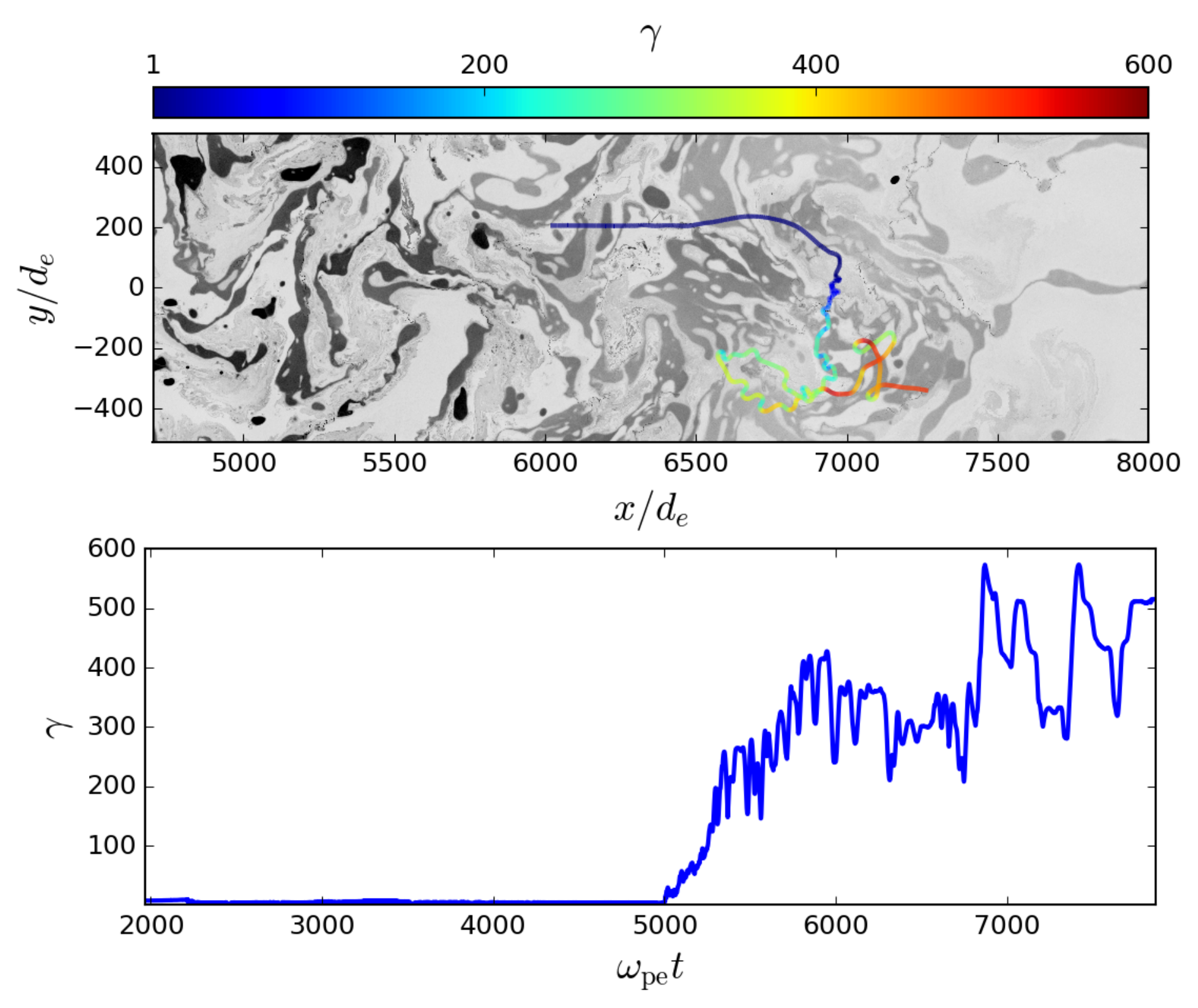}
\caption{Typical high-energy particle trajectories accelerated at the shock front (left panels) and in the turbulent downstream medium (right panels) for $\sigma_0=30$. In the top panels, the Lorentz factor is color-coded along the particle trajectories, itself plotted on top of the plasma density map (gray scale) at the final time of the particle tracking. Bottom panels: Time evolution of the particle Lorentz factor.}
\label{fig_trajectories}
\end{figure*}

Figure~\ref{fig_trajectories} shows two typical high-energy particle trajectories out of a randomly selected sample of $2000$ particles for $\sigma_0=30$, which are meant to illustrate the different acceleration processes at work. Particle~1 shown in the left panels represents the acceleration history of the highest energy particles found in the simulations. As already pointed out in Sect.~\ref{sect_phase_space}, these particles inflate the elongated cavities near the shock front and in the upstream medium. In the early phases ($\omega_{\rm pe}t\lesssim 2500$), the particle is trapped inside the cavity with little acceleration. At $\omega_{\rm pe}t\sim 2500$, the shock front catches up with the particle, which leads to a rapid and uninterrupted acceleration up to at least $\gamma\gtrsim 2000$. During this phase, the particle is trapped in a region that looks like a wake produced behind the cavities where the current layer forms and departs from. The particle trajectory moves back and forth across the equatorial plane where the magnetic field reverses, such that it is well described by the relativistic analog of Speiser orbits \citep{1965JGR....70.4219S, 2012ApJ...746..148C}. As the particle surfs on the wake, its Larmor radius becomes sufficiently large to experience the strong macroscopic bulk-velocity shear, and therefore we associate the acceleration of these particles with the tangential shear-flow acceleration mechanism, which in essence is another form of the Fermi process. In this regime, the energy gain is due to Lorentz-frame transformation as the particle is scattered back and forth across the velocity-shear layer, and is of order $\Delta\gamma/\gamma\sim \Gamma_{\rm s}-1$ after each crossing, where $\Gamma_{\rm s}=1/\sqrt{1-\Delta V^2/c^2}$, and $\Delta V=(V_1-V_2)/(1-V_1 V_2/c^2)$ is the velocity shear sampled by the particle between frames $1$ and $2$ \citep{1990A&A...238..435O, 2004ApJ...617..155R}. In this simulation, the velocity shear is mildly relativistic, $\Delta V/c\sim 0.8$, leading to $\Delta\gamma /\gamma \sim 0.7$. The acceleration proceeds until the particle is kicked out and advected downstream, which plays the role of an escape mechanism.

Particle~2 is more representative of particles accelerated in the turbulent flow further downstream, and therefore of the bulk of the energetic particles in the simulations. It is injected at intermediate latitudes and flows in the laminar magnetized medium between the equatorial and polar current layers without significant energy gain. At $\omega_{\rm pe}t\sim 5000$, the particle is captured by a current layer where it experiences an abrupt acceleration, from $\gamma\sim \Gamma_0$ to $\gamma\sim 200$. We associate this impulsive phenomenon as direct acceleration via relativistic reconnection occurring within the current sheet, which naturally boosts the particle energy to $\gamma\sim \Gamma_0\sigma_0=200$ (e.g., \citealt{2016ApJ...816L...8W}). This event is followed by a much slower stochastic acceleration. At this stage, the particle has reached the turbulent flow where current layers are mixed together and collide at nearly random velocities. This environment favors multiple particle scattering which leads to a stochastic increase or decrease of its energy, but with a net positive gain. In this sense, this process is reminiscent of a second-order Fermi acceleration.

\section{Synchrotron radiation}\label{sect_synchrotron}

\begin{figure}
\centering
\includegraphics[width=\hsize]{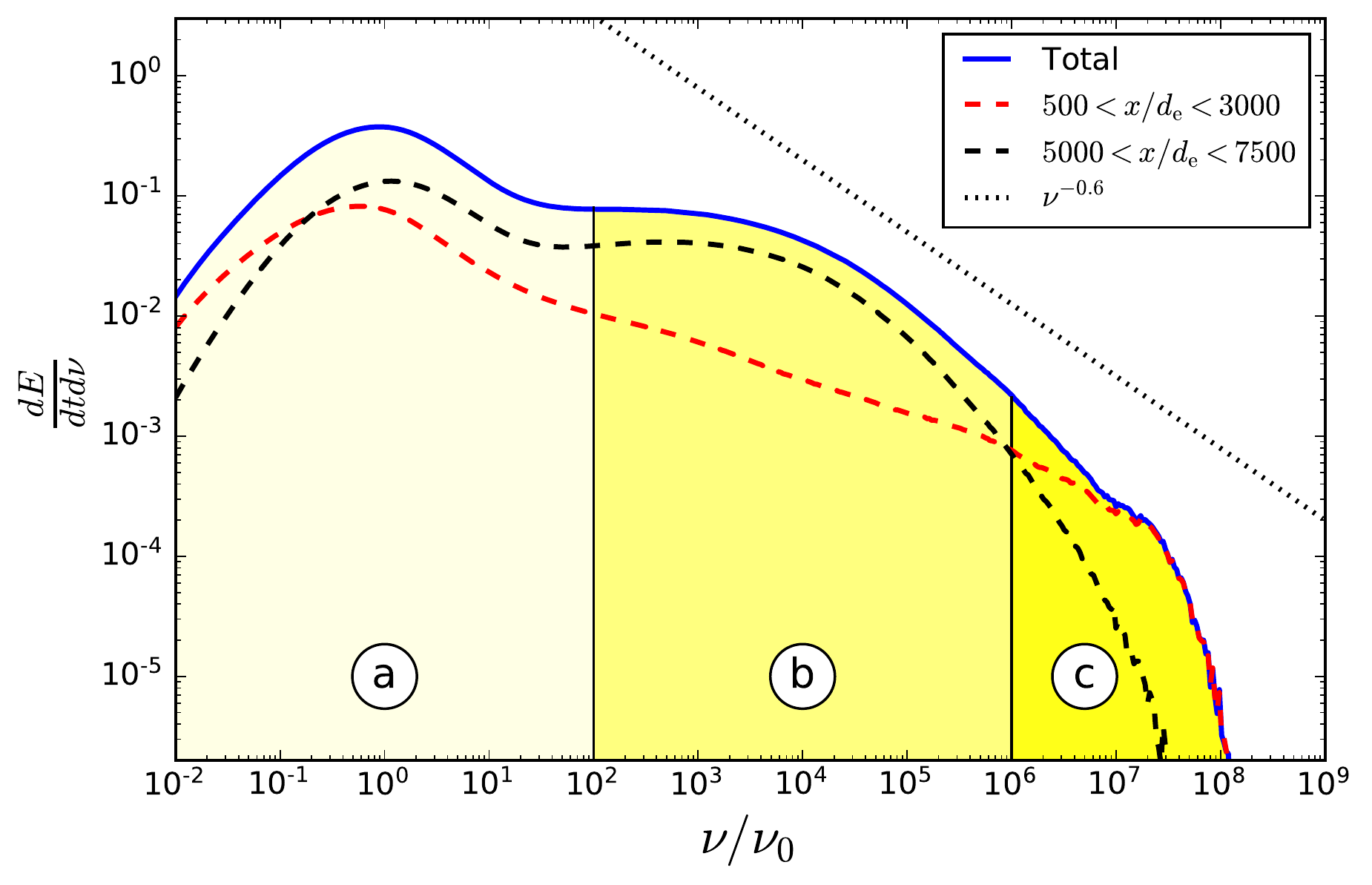}
\caption{Total synchrotron spectrum for $\sigma_0=30$ at $\omega_{\rm pe}t=7840$ (blue solid line). The dashed lines show the spectra emitted near the shock front (red) and in the far-downstream region (black). The dotted line is a pure power law which would be emitted by a $p=2.2$ power-law electron spectrum, such that $dE/dtd\nu\propto \nu^{(-p+1)/2}=\nu^{-0.6}$. The frequency bands labeled \raisebox{.5pt}{\textcircled{a}}, \raisebox{.5pt}{\textcircled{b}}, and \raisebox{.5pt}{\textcircled{c}} refer to Figure~\ref{fig_map_synchrotron}.}
\label{fig_synchrotron}
\end{figure}

\begin{figure*}
\centering
\includegraphics[width=\hsize]{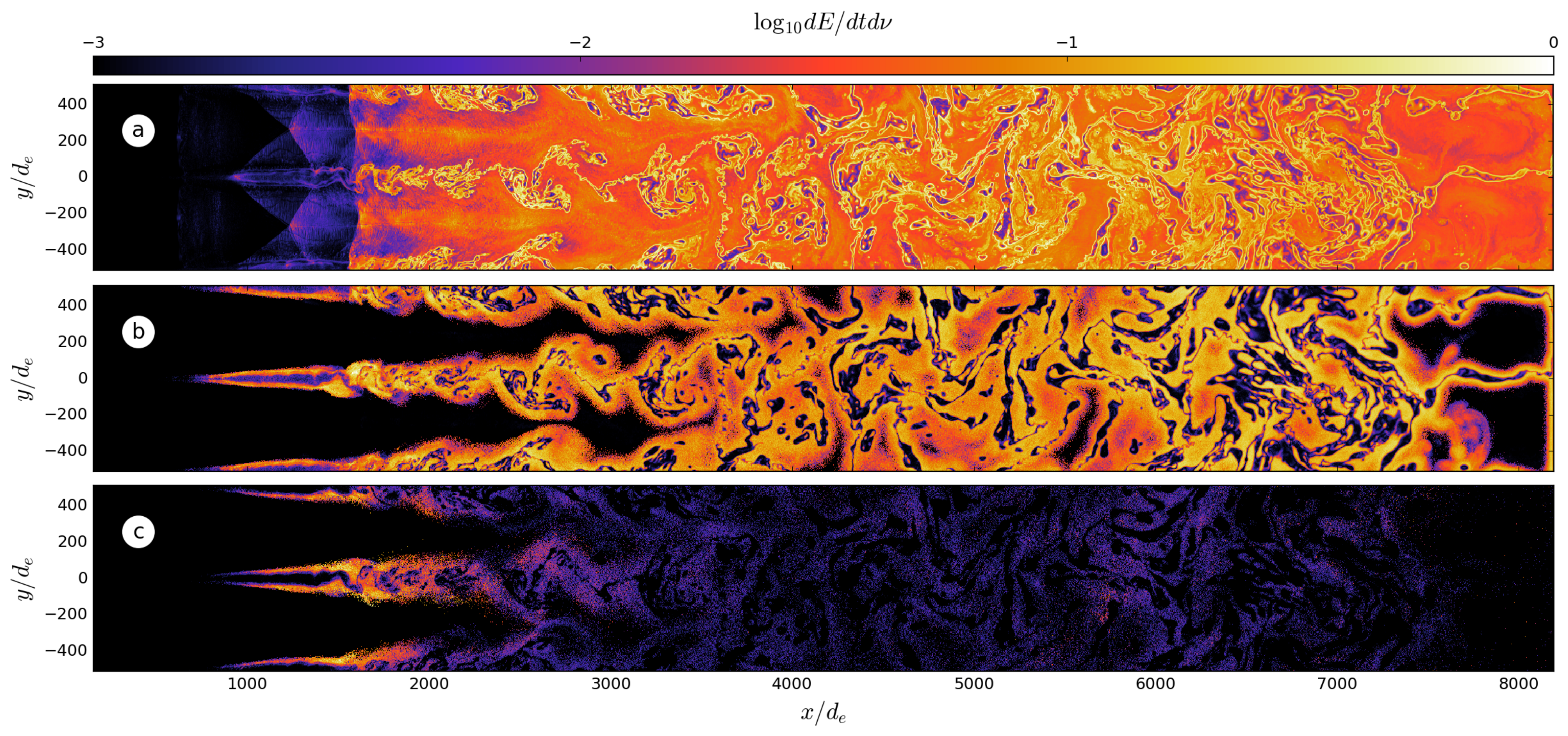}
\caption{Spatial distribution of the synchrotron radiation flux integrated in the frequency bands defined in Figure~\ref{fig_synchrotron}: \raisebox{.5pt}{\textcircled{a}} ($\nu/\nu_0<10^2$), \raisebox{.5pt}{\textcircled{b}} ($10^2<\nu/\nu_0<10^6$), and \raisebox{.5pt}{\textcircled{c}} ($\nu/\nu_0>10^6$).}
\label{fig_map_synchrotron}
\end{figure*}

Figure~\ref{fig_synchrotron} shows the instantaneous synchrotron spectrum emitted by the pairs in the $\sigma_0=30$ simulation at $\omega_{\rm pe}t=7840$, assuming the plasma is optically thin everywhere. To reconstruct the spectrum, we use a delta-function approximation. Each particle emits a single photon radiating away the total power lost by the parent particle, $P_{\rm sync}$, with a frequency set by the synchrotron critical frequency. We recall here that radiative losses are neglected in the simulations for simplicity, and therefore the synchrotron spectrum computed here is not meant to be compared with, for instance, the observed Crab Nebula spectrum, which most likely results from a cooled particle distribution. Instead, our goal here is to characterize the main emission pattern at the shock in different frequency bands. The power spectrum of a single photon is then given by
\begin{equation}
\frac{dE}{dtd\nu}\approx P_{\rm sync}\delta\left(\nu-\nu_{\rm c}\right),
\end{equation}
where
\begin{equation}
P_{\rm sync}=\frac{2}{3}r^2_{\rm e}c\gamma^2 \tilde{B}^2_{\perp},
\end{equation}
$r_{\rm e}=e^2/m_{\rm e}c^2$ is the classical radius of the electron, and
\begin{equation}
\nu_{\rm c}=\frac{3e}{4\pi m_{\rm e}c}\gamma^2\tilde{B}_{\perp},
\end{equation}
is the critical synchrotron frequency. $\tilde{B}_{\perp}$ is the effective perpendicular (to the particle velocity vector) magnetic field in the presence of a strong electric field given by (e.g., \citealt{2016MNRAS.463L..89C})
\begin{equation}
\mathbf{\tilde{B}_{\perp}}=\mathbf{E} + \boldsymbol{\beta}\times\mathbf{B} -\left(\boldsymbol{\beta} \cdot \mathbf{E}\right) \boldsymbol{\beta}.
\end{equation}
Frequencies are normalized by the fiducial synchrotron critical frequency
\begin{equation}
\nu_0=\frac{3e}{4\pi m_{\rm e}c}\Gamma_0^2 B_0.
\end{equation}
The total synchrotron spectrum presents three main features: (i) a low-energy bump centered around $\nu/\nu_0\sim 1$, (ii) a plateau at intermediate frequencies ($dE/dtd\nu\propto \nu^0$, $10^1\lesssim \nu/\nu_0\lesssim 10^4$) followed by, (iii) a steep power-law decline cutting off at $\nu/\nu_0\sim 10^8$. The latter is slightly steeper than the canonical $dE/dtd\nu\propto\nu^{(-p+1)/2}=\nu^{-0.6}$ synchrotron spectrum emitted by a $p=2.2$ power-law electron spectrum immersed into a uniform magnetic field.

The bulk of the synchrotron spectrum originates from the far-downstream region where the flow isotropizes and turbulent reconnection accelerates the particles (see black dashed line in Figure~\ref{fig_synchrotron}). The spectrum emitted in the vicinity of the shock front is composed of the low-energy bump and a hard power-law tail ($dE/dtd\nu\propto\nu^{-0.3}$) spanning over nearly seven orders of magnitude in frequency range. The contribution from the shock front is subdominant at almost all frequencies except at the high-end of the spectrum, $\nu/\nu_0\gtrsim 10^6$ where it rises above the steeper spectrum of the far-downstream region. Figure~\ref{fig_map_synchrotron} shows the spatial distribution of the emitted synchrotron flux integrated in three frequency bands: (i) low-, $\nu/\nu_0<10^2$, (ii) intermediate-, $10^2<\nu/\nu_0<10^6$, and (iii) high-frequency band $\nu/\nu_0>10^6$ labeled as \raisebox{.5pt}{\textcircled{a}}, \raisebox{.5pt}{\textcircled{b}} and \raisebox{.5pt}{\textcircled{c}} respectively. The low-energy flux is rather uniformly distributed in the downstream flow, that is, from the shock front to the back end of the numerical box, from the most magnetized regions to the current layers, but with the notable exception of the shock front cavities. At intermediate frequencies in band \raisebox{.5pt}{\textcircled{b}}, the emission in concentrated within the outer edges of the current layers up to the shock front cavities which are lighting up in this band. Although high-energy particles are also located inside the layers (see Figure~\ref{fig_phase_space}), they do not emit significant radiation because the fields almost vanish in there. Away from the current layers, in smooth magnetized regions, the field is high but energetic particles are absent leading to no flux from these regions. In the high-energy band (\raisebox{.5pt}{\textcircled{c}}), the emission is dominated by the edges of the shock-front cavities piercing through the upstream, as well as by the base of the current sheets in the near downstream medium. The rest of the upstream flow remains dark in all bands as expected because $\tilde{B}_{\perp}\approx 0$.

\section{Discussion and conclusion}\label{sect_conclusion}

Here, we show that the anisotropic nature of the pulsar wind has a dramatic effect on the structure and evolution of the shock front. The usual local plane-parallel approximation does not apply because of the critical role of the global dynamics of the downstream flow, and therefore a latitudinally broad simulation box is essential. A salient feature is the formation of a sharp velocity shear between strongly- and weakly-magnetized regions, which combined with current-driven instabilities leads to strong plasma turbulence in the downstream flow. Current sheets forming in the equatorial plane and at the poles mix and reconnect leading to efficient nonthermal particle acceleration. The efficiency of particle acceleration increases with $\sigma$, which is similar to relativistic reconnection but opposite to Fermi acceleration in uniform shocks. Turbulent reconnection leads to a power-law electron spectrum with a slope that hardens with $\sigma$, as expected from recent studies of relativistic reconnection \citep{2014ApJ...783L..21S, 2016ApJ...816L...8W} and kinetic turbulence \citep{2017PhRvL.118e5103Z, 2018ApJ...867L..18Z}. Scaled up to the Crab Nebula, with $dN/d\gamma\propto \gamma^{-2.2}$, the injected (uncooled) X-ray electron spectrum is consistent with the high-$\sigma$ shock solution $\sigma_0=30$, or a mean upstream magnetization $\bar{\sigma}_0\approx 5$. This result is compatible with global 3D MHD models which advocate high-$\sigma$ solutions to explain the morphology of the Crab Nebula \citep{2013MNRAS.431L..48P, 2014MNRAS.438..278P}. The spectrum extends from $\Gamma_0$ to $\sim\Gamma_0\sigma_0$ beyond which it steepens significantly and where another component takes over and extends the total spectrum to even higher energies. It is tempting to connect this result with the mysterious high-energy break in the Crab Nebula spectrum, where the electron spectral index decreases by $\sim 0.5$ \citep{2010A&A...523A...2M}.

The high-energy component originates from another robust and extraordinary feature of anisotropic magnetized shocks, which is the formation of elongated cavities at the base of the polar and equatorial regions drilling through the upstream medium. These structures are low-field, low-density regions moving with mildly relativistic speeds against the incoming flow, and their sizes continuously grow with time. They are inflated by the highest energy particles in the box, which follow relativistic Speiser orbits. These special trajectories, typically found in reconnection layers, are captured by the midplane where the magnetic field polarity reverses. This trapping mechanism provides stability to the cavities themselves, whose sizes constantly adjust to the particle Larmor radius. These particles are energized via shear-flow acceleration at the interface between the cavities and the incoming flow. This component alone explains the highest-energy part of the particle spectrum. The maximum particle energy increases nearly linearly with time, $\gamma_{\rm max}\propto t$, in contrast to Weibel-dominated shock acceleration where $\gamma_{\rm max}\propto\sqrt{t}$ \citep{2013ApJ...771...54S, 2018MNRAS.477.5238P}, meaning that the acceleration process is very efficient, close to the Bohm limit. Another important difference with Weibel-dominated shock acceleration is that the particle spectrum does not show signs of saturation, the maximum energy grows steadily until the end of the simulations. Presumably, the particle energy will be limited by the transverse size of the shock or by radiative losses such as those in the Crab Nebula where the electron maximum energy is limited by the synchrotron burn-off limit \citep{1983MNRAS.205..593G, 1996ApJ...457..253D}. These cavities also have the peculiarity to preferentially accelerate one sign of charge, electrons in the equatorial region, positrons (ions) at the poles, and vice-versa if $\boldsymbol{\Omega}\cdot\mathbf{B}<0$, as also reported by \citet{2018ApJ...863...18G}.

The modeling of synchrotron radiation indicates that while the bulk of the emission is produced quasi-isotropically in the downstream region, the high-end of the synchrotron spectrum is concentrated within the edges and the wakes of the shock front cavities where both the field and the particle energy are the highest. Translated in terms of the Crab Nebula features, we predict that the 100~MeV emission is preferably localized at the inner ring, and specifically on the side receding away from us as the emission in the cavities is Doppler boosted in the direction opposite to the incoming pulsar wind. There might also be a weaker contribution from the base of the counter jet because of the smaller volume involved in comparison with the equatorial region. The wake behind the cavities is highly dynamical. Due to the kink instability, the high-energy beam sweeps a wide angular range. Therefore, gamma-ray flares at the high end of the synchrotron spectrum come out as a natural consequence of particle acceleration at the pulsar wind termination shock, and more generally in any relativistic, magnetized, and anisotropic shocks with a possible application to the hotspots of relativistic jets and gamma-ray bursts. The mildly relativistic bulk motion of the backflow, with $\Gamma\sim 3-4$, would naturally push the radiation energy above the rest-frame 160~MeV synchrotron burn-off limit, a persistent feature of the Crab Nebula gamma-ray flares \citep{2011Sci...331..736T, 2011Sci...331..739A}. Although these results suggest a connection between particle acceleration in the cavities and gamma-ray flares, more work is needed for a solid conclusion.

An obvious limitation of the proposed model is its Cartesian plane-parallel geometry. Any curvature of the shock front and the radial expansion of the downstream flow are therefore neglected in this work. A more realistic configuration with a spherical geometry and a finite curvature of the shock front would be the logical next step to confirm our findings. This would also break the symmetry between the poles and the equator, which play nearly identical roles in this study. In particular, the cavity in the equator may then have a dominant contribution in the acceleration (charge asymmetry) and radiation pattern (gamma-ray flares) because of its larger volume in comparison with the poles. An extension to 3D simulations would also be desirable to fully capture magnetic reconnection that occurs here in the out-of-plane direction and therefore some of the important features of reconnection are missing here (e.g., tearing mode and plasmoid formation). Three-dimensional simulations can also capture possible departures from axisymmetry, which may explain for instance the knotty nature of the inner ring in the Crab Nebula. Synchrotron cooling could play an important role in the dynamics of the shock front cavities where it is most severe. Although they are probably highly subdominant in number compared with pairs, ions are most likely present in the wind and in the nebula. If the efficient particle acceleration mechanism revealed in this work also applies to them, pulsar wind nebulae could be an important source of the Galactic cosmic-ray population. Dissipation of the oscillating current sheet in the pulsar wind may lead to efficient particle acceleration ahead of the termination shock \citep{2017A&A...607A.134C}. An excess of energetic particles in the equatorial regions may affect the shock dynamics in return. The exploration of all of the above effects provides a wide array of possible future investigations.

\begin{acknowledgements}
We are grateful to the anonymous referee for his/her very supportive report. This work has been supported by the Programme National des Hautes \'Energies of CNRS/INSU and CNES. We acknowledge PRACE and GENCI (allocations A0050407669 and A0070407669) for awarding us access to Joliot-Curie at GENCI@CEA, France.
\end{acknowledgements}

\bibliographystyle{aa}
\bibliography{shock_pwn}

\begin{thebibliography}{52}
\expandafter\ifx\csname natexlab\endcsname\relax\def\natexlab#1{#1}\fi

\bibitem[{{Abdo} {et~al.}(2011){Abdo}, {Ackermann}, {Ajello}, {Allafort},
  {Baldini}, {Ballet}, {Barbiellini}, {Bastieri}, {Bechtol}, {Bellazzini},
  {Berenji}, {Blandford}, {Bloom}, {Bonamente}, {Borgland}, {Bouvier}, {Brand
  t}, {Bregeon}, {Brez}, {Brigida}, {Bruel}, {Buehler}, {Buson}, {Caliandro},
  {Cameron}, {Cannon}, {Caraveo}, {Casand jian}, {{\c{C}}elik}, {Charles},
  {Chekhtman}, {Cheung}, {Chiang}, {Ciprini}, {Claus}, {Cohen-Tanugi},
  {Costamante}, {Cutini}, {D'Ammando}, {Dermer}, {de Angelis}, {de Luca}, {de
  Palma}, {Digel}, {do Couto e Silva}, {Drell}, {Drlica-Wagner}, {Dubois},
  {Dumora}, {Favuzzi}, {Fegan}, {Ferrara}, {Focke}, {Fortin}, {Frailis},
  {Fukazawa}, {Funk}, {Fusco}, {Gargano}, {Gasparrini}, {Gehrels}, {Germani},
  {Giglietto}, {Giordano}, {Giroletti}, {Glanzman}, {Godfrey}, {Grenier},
  {Grondin}, {Grove}, {Guiriec}, {Hadasch}, {Hanabata}, {Harding}, {Hayashi},
  {Hayashida}, {Hays}, {Horan}, {Itoh}, {J{\'o}hannesson}, {Johnson},
  {Johnson}, {Khangulyan}, {Kamae}, {Katagiri}, {Kataoka}, {Kerr},
  {Kn{\"o}dlseder}, {Kuss}, {Lande}, {Latronico}, {Lee}, {Lemoine-Goumard},
  {Longo}, {Loparco}, {Lubrano}, {Madejski}, {Makeev}, {Marelli}, {Mazziotta},
  {McEnery}, {Michelson}, {Mitthumsiri}, {Mizuno}, {Moiseev}, {Monte},
  {Monzani}, {Morselli}, {Moskalenko}, {Murgia}, {Nakamori}, {Naumann-Godo},
  {Nolan}, {Norris}, {Nuss}, {Ohsugi}, {Okumura}, {Omodei}, {Ormes}, {Ozaki},
  {Paneque}, {Parent}, {Pelassa}, {Pepe}, {Pesce-Rollins}, {Pierbattista},
  {Piron}, {Porter}, {Rain{\`o}}, {Rando}, {Ray}, {Razzano}, {Reimer},
  {Reimer}, {Reposeur}, {Ritz}, {Romani}, {Sadrozinski}, {Sanchez},
  {Parkinson}, {Scargle}, {Schalk}, {Sgr{\`o}}, {Siskind}, {Smith}, {Spand re},
  {Spinelli}, {Strickman}, {Suson}, {Takahashi}, {Takahashi}, {Tanaka},
  {Thayer}, {Thompson}, {Tibaldo}, {Torres}, {Tosti}, {Tramacere}, {Troja},
  {Uchiyama}, {Vandenbroucke}, {Vasileiou}, {Vianello}, {Vitale}, {Wang},
  {Wood}, {Yang}, \& {Ziegler}}]{2011Sci...331..739A}
{Abdo}, A.~A., {Ackermann}, M., {Ajello}, M., {et~al.} 2011, Science, 331, 739

\bibitem[{{Achterberg} {et~al.}(2001){Achterberg}, {Gallant}, {Kirk}, \&
  {Guthmann}}]{2001MNRAS.328..393A}
{Achterberg}, A., {Gallant}, Y.~A., {Kirk}, J.~G., \& {Guthmann}, A.~W. 2001,
  \mnras, 328, 393

\bibitem[{{Amato}(2020)}]{2020arXiv200104442A}
{Amato}, E. 2020, arXiv e-prints, arXiv:2001.04442

\bibitem[{{Amato} \& {Arons}(2006)}]{2006ApJ...653..325A}
{Amato}, E. \& {Arons}, J. 2006, \apj, 653, 325

\bibitem[{{Bednarz} \& {Ostrowski}(1998)}]{1998PhRvL..80.3911B}
{Bednarz}, J. \& {Ostrowski}, M. 1998, \prl, 80, 3911

\bibitem[{{Begelman}(1998)}]{1998ApJ...493..291B}
{Begelman}, M.~C. 1998, \apj, 493, 291

\bibitem[{{Begelman} \& {Kirk}(1990)}]{1990ApJ...353...66B}
{Begelman}, M.~C. \& {Kirk}, J.~G. 1990, \apj, 353, 66

\bibitem[{{Bogovalov}(1999)}]{1999A&A...349.1017B}
{Bogovalov}, S.~V. 1999, \aap, 349, 1017

\bibitem[{{Cerutti} {et~al.}(2016){Cerutti}, {Mortier}, \&
  {Philippov}}]{2016MNRAS.463L..89C}
{Cerutti}, B., {Mortier}, J., \& {Philippov}, A.~A. 2016, \mnras, 463, L89

\bibitem[{{Cerutti} \& {Philippov}(2017)}]{2017A&A...607A.134C}
{Cerutti}, B. \& {Philippov}, A.~A. 2017, \aap, 607, A134

\bibitem[{{Cerutti} {et~al.}(2012){Cerutti}, {Uzdensky}, \&
  {Begelman}}]{2012ApJ...746..148C}
{Cerutti}, B., {Uzdensky}, D.~A., \& {Begelman}, M.~C. 2012, \apj, 746, 148

\bibitem[{{Cerutti} \& {Werner}(2019)}]{2019ascl.soft11012C}
{Cerutti}, B. \& {Werner}, G. 2019, {Zeltron: Explicit 3D relativistic
  electromagnetic Particle-In-Cell code}

\bibitem[{{Cerutti} {et~al.}(2013){Cerutti}, {Werner}, {Uzdensky}, \&
  {Begelman}}]{2013ApJ...770..147C}
{Cerutti}, B., {Werner}, G.~R., {Uzdensky}, D.~A., \& {Begelman}, M.~C. 2013,
  \apj, 770, 147

\bibitem[{{Coroniti}(1990)}]{1990ApJ...349..538C}
{Coroniti}, F.~V. 1990, \apj, 349, 538

\bibitem[{{de Jager} {et~al.}(1996){de Jager}, {Harding}, {Michelson}, {Nel},
  {Nolan}, {Sreekumar}, \& {Thompson}}]{1996ApJ...457..253D}
{de Jager}, O.~C., {Harding}, A.~K., {Michelson}, P.~F., {et~al.} 1996, \apj,
  457, 253

\bibitem[{{Demidem} {et~al.}(2018){Demidem}, {Lemoine}, \&
  {Casse}}]{2018MNRAS.475.2713D}
{Demidem}, C., {Lemoine}, M., \& {Casse}, F. 2018, \mnras, 475, 2713

\bibitem[{{Gallant} {et~al.}(1992){Gallant}, {Hoshino}, {Langdon}, {Arons}, \&
  {Max}}]{1992ApJ...391...73G}
{Gallant}, Y.~A., {Hoshino}, M., {Langdon}, A.~B., {Arons}, J., \& {Max}, C.~E.
  1992, \apj, 391, 73

\bibitem[{{Giacinti} \& {Kirk}(2018)}]{2018ApJ...863...18G}
{Giacinti}, G. \& {Kirk}, J.~G. 2018, \apj, 863, 18

\bibitem[{{Greenwood} {et~al.}(2004){Greenwood}, {Cartwright}, {Luginsland },
  \& {Baca}}]{2004JCoPh.201..665G}
{Greenwood}, A.~D., {Cartwright}, K.~L., {Luginsland }, J.~W., \& {Baca}, E.~A.
  2004, Journal of Computational Physics, 201, 665

\bibitem[{{Guilbert} {et~al.}(1983){Guilbert}, {Fabian}, \&
  {Rees}}]{1983MNRAS.205..593G}
{Guilbert}, P.~W., {Fabian}, A.~C., \& {Rees}, M.~J. 1983, \mnras, 205, 593

\bibitem[{{Hoshino} {et~al.}(1992){Hoshino}, {Arons}, {Gallant}, \&
  {Langdon}}]{1992ApJ...390..454H}
{Hoshino}, M., {Arons}, J., {Gallant}, Y.~A., \& {Langdon}, A.~B. 1992, \apj,
  390, 454

\bibitem[{{Kennel} \& {Coroniti}(1984)}]{1984ApJ...283..710K}
{Kennel}, C.~F. \& {Coroniti}, F.~V. 1984, \apj, 283, 710

\bibitem[{{Keshet} {et~al.}(2009){Keshet}, {Katz}, {Spitkovsky}, \&
  {Waxman}}]{2009ApJ...693L.127K}
{Keshet}, U., {Katz}, B., {Spitkovsky}, A., \& {Waxman}, E. 2009, \apjl, 693,
  L127

\bibitem[{{Kirk} {et~al.}(2000){Kirk}, {Guthmann}, {Gallant}, \&
  {Achterberg}}]{2000ApJ...542..235K}
{Kirk}, J.~G., {Guthmann}, A.~W., {Gallant}, Y.~A., \& {Achterberg}, A. 2000,
  \apj, 542, 235

\bibitem[{{Kirk} {et~al.}(2009){Kirk}, {Lyubarsky}, \&
  {Petri}}]{2009ASSL..357..421K}
{Kirk}, J.~G., {Lyubarsky}, Y., \& {Petri}, J. 2009, Astrophysics and Space
  Science Library, Vol. 357, {The Theory of Pulsar Winds and Nebulae}, ed.
  W.~{Becker}, 421

\bibitem[{{Komissarov}(2013)}]{2013MNRAS.428.2459K}
{Komissarov}, S.~S. 2013, \mnras, 428, 2459

\bibitem[{{Langdon} {et~al.}(1988){Langdon}, {Arons}, \&
  {Max}}]{1988PhRvL..61..779L}
{Langdon}, A.~B., {Arons}, J., \& {Max}, C.~E. 1988, \prl, 61, 779

\bibitem[{{Lemoine}(2016)}]{2016JPlPh..82d6301L}
{Lemoine}, M. 2016, Journal of Plasma Physics, 82, 635820401

\bibitem[{{Lemoine} {et~al.}(2019){Lemoine}, {Gremillet}, {Pelletier}, \&
  {Vanthieghem}}]{2019PhRvL.123c5101L}
{Lemoine}, M., {Gremillet}, L., {Pelletier}, G., \& {Vanthieghem}, A. 2019,
  \prl, 123, 035101

\bibitem[{{Lemoine} \& {Pelletier}(2010)}]{2010MNRAS.402..321L}
{Lemoine}, M. \& {Pelletier}, G. 2010, \mnras, 402, 321

\bibitem[{{Lyubarsky}(2003)}]{2003MNRAS.345..153L}
{Lyubarsky}, Y.~E. 2003, \mnras, 345, 153

\bibitem[{{Medvedev} \& {Loeb}(1999)}]{1999ApJ...526..697M}
{Medvedev}, M.~V. \& {Loeb}, A. 1999, \apj, 526, 697

\bibitem[{{Meyer} {et~al.}(2010){Meyer}, {Horns}, \&
  {Zechlin}}]{2010A&A...523A...2M}
{Meyer}, M., {Horns}, D., \& {Zechlin}, H.~S. 2010, \aap, 523, A2

\bibitem[{{Michel}(1973)}]{1973ApJ...180L.133M}
{Michel}, F.~C. 1973, \apjl, 180, L133

\bibitem[{{Olmi} {et~al.}(2015){Olmi}, {Del Zanna}, {Amato}, \&
  {Bucciantini}}]{2015MNRAS.449.3149O}
{Olmi}, B., {Del Zanna}, L., {Amato}, E., \& {Bucciantini}, N. 2015, \mnras,
  449, 3149

\bibitem[{{Ostrowski}(1990)}]{1990A&A...238..435O}
{Ostrowski}, M. 1990, \aap, 238, 435

\bibitem[{{Pelletier} {et~al.}(2017){Pelletier}, {Bykov}, {Ellison}, \&
  {Lemoine}}]{2017SSRv..207..319P}
{Pelletier}, G., {Bykov}, A., {Ellison}, D., \& {Lemoine}, M. 2017, \ssr, 207,
  319

\bibitem[{{P{\'e}tri} \& {Lyubarsky}(2007)}]{2007A&A...473..683P}
{P{\'e}tri}, J. \& {Lyubarsky}, Y. 2007, \aap, 473, 683

\bibitem[{{Plotnikov} {et~al.}(2018){Plotnikov}, {Grassi}, \&
  {Grech}}]{2018MNRAS.477.5238P}
{Plotnikov}, I., {Grassi}, A., \& {Grech}, M. 2018, \mnras, 477, 5238

\bibitem[{{Porth} {et~al.}(2013){Porth}, {Komissarov}, \&
  {Keppens}}]{2013MNRAS.431L..48P}
{Porth}, O., {Komissarov}, S.~S., \& {Keppens}, R. 2013, \mnras, 431, L48

\bibitem[{{Porth} {et~al.}(2014){Porth}, {Komissarov}, \&
  {Keppens}}]{2014MNRAS.438..278P}
{Porth}, O., {Komissarov}, S.~S., \& {Keppens}, R. 2014, \mnras, 438, 278

\bibitem[{{Rees} \& {Gunn}(1974)}]{1974MNRAS.167....1R}
{Rees}, M.~J. \& {Gunn}, J.~E. 1974, \mnras, 167, 1

\bibitem[{{Rieger} \& {Duffy}(2004)}]{2004ApJ...617..155R}
{Rieger}, F.~M. \& {Duffy}, P. 2004, \apj, 617, 155

\bibitem[{{Sironi} \& {Spitkovsky}(2011)}]{2011ApJ...741...39S}
{Sironi}, L. \& {Spitkovsky}, A. 2011, \apj, 741, 39

\bibitem[{{Sironi} \& {Spitkovsky}(2014)}]{2014ApJ...783L..21S}
{Sironi}, L. \& {Spitkovsky}, A. 2014, \apjl, 783, L21

\bibitem[{{Sironi} {et~al.}(2013){Sironi}, {Spitkovsky}, \&
  {Arons}}]{2013ApJ...771...54S}
{Sironi}, L., {Spitkovsky}, A., \& {Arons}, J. 2013, \apj, 771, 54

\bibitem[{{Speiser}(1965)}]{1965JGR....70.4219S}
{Speiser}, T.~W. 1965, \jgr, 70, 4219

\bibitem[{{Spitkovsky}(2008)}]{2008ApJ...682L...5S}
{Spitkovsky}, A. 2008, \apjl, 682, L5

\bibitem[{{Tavani} {et~al.}(2011){Tavani}, {Bulgarelli}, {Vittorini},
  {Pellizzoni}, {Striani}, {Caraveo}, {Weisskopf}, {Tennant}, {Pucella},
  {Trois}, {Costa}, {Evangelista}, {Pittori}, {Verrecchia}, {Del Monte},
  {Campana}, {Pilia}, {De Luca}, {Donnarumma}, {Horns}, {Ferrigno}, {Heinke},
  {Trifoglio}, {Gianotti}, {Vercellone}, {Argan}, {Barbiellini}, {Cattaneo},
  {Chen}, {Contessi}, {D'Ammand o}, {DeParis}, {Di Cocco}, {Di Persio},
  {Feroci}, {Ferrari}, {Galli}, {Giuliani}, {Giusti}, {Labanti}, {Lapshov},
  {Lazzarotto}, {Lipari}, {Longo}, {Fuschino}, {Marisaldi}, {Mereghetti},
  {Morelli}, {Moretti}, {Morselli}, {Pacciani}, {Perotti}, {Piano}, {Picozza},
  {Prest}, {Rapisarda}, {Rappoldi}, {Rubini}, {Sabatini}, {Soffitta},
  {Vallazza}, {Zambra}, {Zanello}, {Lucarelli}, {Santolamazza}, {Giommi},
  {Salotti}, \& {Bignami}}]{2011Sci...331..736T}
{Tavani}, M., {Bulgarelli}, A., {Vittorini}, V., {et~al.} 2011, Science, 331,
  736

\bibitem[{{Werner} {et~al.}(2016){Werner}, {Uzdensky}, {Cerutti}, {Nalewajko},
  \& {Begelman}}]{2016ApJ...816L...8W}
{Werner}, G.~R., {Uzdensky}, D.~A., {Cerutti}, B., {Nalewajko}, K., \&
  {Begelman}, M.~C. 2016, \apjl, 816, L8

\bibitem[{{Zhdankin} {et~al.}(2018){Zhdankin}, {Uzdensky}, {Werner}, \&
  {Begelman}}]{2018ApJ...867L..18Z}
{Zhdankin}, V., {Uzdensky}, D.~A., {Werner}, G.~R., \& {Begelman}, M.~C. 2018,
  \apjl, 867, L18

\bibitem[{{Zhdankin} {et~al.}(2017){Zhdankin}, {Werner}, {Uzdensky}, \&
  {Begelman}}]{2017PhRvL.118e5103Z}
{Zhdankin}, V., {Werner}, G.~R., {Uzdensky}, D.~A., \& {Begelman}, M.~C. 2017,
  \prl, 118, 055103

\end{thebibliography}

\end{document}